%% file: paper-elsarticle.tex
\documentclass[11.5pt, final,1p,times, review]{elsarticle}

\usepackage[ruled,vlined,linesnumbered]{algorithm2e}
\usepackage{hyperref}

\usepackage{array}

\usepackage[misc,geometry]{ifsym}
\usepackage{microtype}
\usepackage[dvipsnames,table,xcdraw]{xcolor}
\usepackage{paralist}

\usepackage[utf8]{inputenc}
\usepackage[T1]{fontenc}
\usepackage{wrapfig}
\usepackage[caption = false]{subfig}
\usepackage{subfloat}
\usepackage{booktabs}
\usepackage{caption}
\usepackage{amssymb, amsfonts}
\usepackage{amsmath}
\usepackage{amsopn}
\usepackage{amsthm}
\usepackage{tikz, tikz-qtree}
\usepackage{url}
\usepackage{balance}

\usepackage{multirow}
\usepackage{relsize}
\usepackage{caption}
\usepackage{pdflscape}
\usepackage{enumitem}
\usepackage{placeins}
\usepackage[normalem]{ulem}
\usepackage{marginnote}
\usepackage{todonotes}

\usepackage{soul}

\usepackage{etoolbox}
\makeatletter
\patchcmd{\SOUL@ulunderline}{\dimen@}{\SOUL@dimen}{}{}
\patchcmd{\SOUL@ulunderline}{\dimen@}{\SOUL@dimen}{}{}
\patchcmd{\SOUL@ulunderline}{\dimen@}{\SOUL@dimen}{}{}
\newdimen\SOUL@dimen
\makeatother

\usetikzlibrary{shapes}
\usetikzlibrary{matrix}
\usetikzlibrary{tikzmark}
\usetikzlibrary{decorations.pathreplacing}
\usetikzlibrary{arrows.meta}

\sethlcolor{Yellow!20}

\setitemize{noitemsep,topsep=0pt,parsep=0pt,partopsep=0pt}

\newtheorem{definition}{Definition}[section]

\newtheorem{property}{Property}
\newtheorem{proof}{Proof}

\begin{document}

\begin{frontmatter}



\title{Mining a Minimal Set of Behavioral Patterns using Incremental
Evaluation}


\author{Mehdi Acheli}{}

\affiliation [label1]{organization={Paris Dauphine - PSL University, LAMSADE UMR 7243 CNRS},
            addressline={},
            city={Paris},
            postcode={75016},
            state={},
            country={France}}

\author{Daniela Grigori}{}

\author{Matthias Weidlich}{}

\affiliation [label2]{organization={Humboldt-Universitat zu Berlin},
            addressline={},
            city={Berlin},
            postcode={},
            state={},
            country={Germany}}

\begin{abstract}
Process mining provides methods to analyse event logs generated by
information systems during the execution of processes. It thereby supports the
design, validation, and execution of processes in domains ranging from
healthcare, through manufacturing, to e-commerce.
To explore the regularities of flexible processes that show a
large behavioral variability, it was suggested to mine recurrent behavioral
patterns that jointly describe the underlying process. Existing approaches
to behavioral pattern mining, however, suffer from two limitations. First,
they show limited scalability as incremental computation is incorporated only
in the generation of pattern candidates, but not in the evaluation of their
quality. Second, process analysis based on mined patterns shows
limited effectiveness due to an overwhelmingly large number of patterns obtained
in practical application scenarios, many of which are redundant. In this paper, we address these limitations to facilitate the analysis of
complex, flexible processes based on behavioral
patterns. Specifically, we improve COBPAM, our initial behavioral pattern mining algorithm, by an
incremental procedure to evaluate the quality of pattern candidates, optimizing thereby its efficiency. Targetting
a more effective use of the resulting patterns, we further propose pruning
strategies for redundant patterns and show how relations between the remaining
patterns are extracted and visualized to provide process insights. Our
experiments with
diverse real-world datasets indicate a considerable reduction of the runtime
needed for pattern mining, while a qualitative assessment highlights how
relations between patterns guide the analysis of the underlying process.
\end{abstract}




\end{frontmatter}


\input{introduction.tex}

\input{related_work.tex}

\input{background.tex}

\input{alignment_growth.tex}

\section{Post-processing}
\label{sec:postpro}

The number of behavioral patterns returned by COBPAM may be overwhelming. In order to cater for this issue, we include further processing steps at the end of its execution. The idea is to avoid outputting trees that can be deduced from others and trees that are equivalent.  The notion of generalized maximality, introduced in section \autoref{sec:genMax} captures if one pattern can be inferred from another one. In \autoref{sec:equiv} we present definitions of equivalent trees.
In order to further improve the understandability of the resulting set of patterns, we propose in section \ref{sec:viz} a graph-based visualisation of patterns and their relationships. 

\subsection{Generalized Maximality}\label{sec:genMax}

Considering all patterns of at most depth i, a pattern P is maximal, if its behavior is not included in that of another pattern P' of depth smaller or equal to $i$, i.e., P is not the regular seed of P'.

In the following, we define a notion of generalized maximality based on other types of seeds.  We then define pruning techniques to remove the redundant patterns in the output of the discovery algorithm to ensure that the resulting set is minimal.

\subsubsection{Alternative Seeds}

The goal is to eliminate all the patterns in the output that can be inferred from the others.
We give in \autoref{fig:exAltSeeds} an example of a tree $T$ with \emph{its} two seeds $T_1$ and $T_2$, respecting the definition \ref{Comb_leaf}. However, we can also see two other trees, that can be combined to get $T$: $T_3$ and $T_4$. These  differ in only one leaf just like the regular seeds albeit that leaf is not at a position of a potential combination leaf. (Restrictions on the position of combination leaf in the definition \ref{Comb_leaf} are imposed to ensure the unicity of the seeds and the efficiency of the combination-based pattern discovery algorithm \cite{acheli2019efficient}). 
$T_3$ and $T_4$, called alternative seeds,  could also be inferred from $T$. The generalized maximality notion captures the  redundancy of alternative seeds.

\begin{figure}[h]
	\centering

	\subfloat[]{\centering
		\begin{tabular}{c}
			\begin{tikzpicture}
			\tikzset{level 1/.style={level distance=24pt}}
			\tikzset{level 2/.style={level distance=20pt}}
			\tikzset{level 3+/.style={level distance=20pt}}
			\Tree [.\textit{seq} [.\textit{seq} [.\textcolor{blue}{a} ] [.\textcolor{blue}{b} ] ] [.\textit{seq} [.\textcolor{red}{c} ] [.\textcolor{red}{d} ] ] ]
			\end{tikzpicture}
		\end{tabular}
	}\\$\swarrow$ \quad \quad \quad $\searrow$\\
	\subfloat[]{\centering
		\begin{tabular}{c}
			\begin{tikzpicture}
			\tikzset{level 1/.style={level distance=24pt}}
			\tikzset{level 2/.style={level distance=20pt}}
			\tikzset{level 3+/.style={level distance=20pt}}
			\Tree [.\textit{seq} [.\textit{seq} [.a ] [.b ] ] [.\textcolor{red}{c} ] ]
			\end{tikzpicture}
		\end{tabular}
	}
		\subfloat[]{\centering
		\begin{tabular}{c}
			\begin{tikzpicture}
			\tikzset{level 1/.style={level distance=24pt}}
			\tikzset{level 2/.style={level distance=20pt}}
			\tikzset{level 3+/.style={level distance=20pt}}
			\Tree [.\textit{seq} [.\textit{seq} [.a ] [.b ] ] [.\textcolor{red}{d} ] ]
			\end{tikzpicture}
		\end{tabular}
	}\quad\quad\quad
	\subfloat[]{\centering
		\begin{tabular}{c}
			\begin{tikzpicture}
			\tikzset{level 1/.style={level distance=24pt}}
			\tikzset{level 2/.style={level distance=20pt}}
			\tikzset{level 3+/.style={level distance=20pt}}
			\Tree [.\textit{seq} [.\textcolor{blue}{a} ] [.\textit{seq} [.c ] [.d ] ] ]
			\end{tikzpicture}
		\end{tabular}
	}
		\subfloat[]{\centering
		\begin{tabular}{c}
			\begin{tikzpicture}
			\tikzset{level 1/.style={level distance=24pt}}
			\tikzset{level 2/.style={level distance=20pt}}
			\tikzset{level 3+/.style={level distance=20pt}}
			\Tree [.\textit{seq} [.\textcolor{blue}{b} ] [.\textit{seq} [.c ] [.d ] ] ]
			\end{tikzpicture}
		\end{tabular}
	}

	\vspace{-.5em}
	\caption{A tree $T$ (a), its seeds: $T_1$ (b) and $T_2$ (c) and two alternative seeds: $T_3$ (d) and $T_4$ (e)}
	\label{fig:exAltSeeds}
	\vspace{-1em}
\end{figure}

In the following, we  give a formalization of the alternative seeds.
\begin{definition}[Alternative seeds]
Given a process tree $P$, we define a function, $f$, that maps $P$ to a set of trees.

\begin{itemize}
    \item if $P = a$ with $a \in A$, an activity, then $f(P) = \{a\}$.
    \item if $P = x(P_1, P_2)$ where $x$ is a constraining operator and $P_1$, $P_2$ children subtrees, $f(P)$ is given as the union of:
    \begin{itemize}
        \item $f(P_1)$.
        \item $f(P_2)$.
        \item the set: $\{x(P'_1, P'_2) \mid P'_1 \in f(P_1), P'_2 \in f(P_2)\}$.
    \end{itemize}
    \item if $P = xor(P_1, P_2)$ with $P_1$, $P_2$ children subtrees, then $f(P) = \{xor(P'_1, P'_2) \mid P'_1 \in f(P_1), P'_2 \in f(P_2)\}$.

\end{itemize}

For a process tree of depth $n \in \mathbb{N}^*$, the elements in $f(P) \setminus \{P\}$ are its seeds (regular and alternative seeds). In particular, alternative seeds are defined for trees of at least depth two where the root is a constraining operator. If we consider such a tree $P$ then its alternative seeds are the elements of $f(P) \setminus \{P\}$ minus the regular seeds defined through the combination operation (see ~\cite{acheli2019efficient}). The depth condition is justified by the absence of seeds for depth zero and the absence of alternative seeds (only regular) for depth one.
\end{definition}

A generalized monotonicity property states that if a process tree $P$ is frequent then all its alternative seeds are frequent too\footnote{The proof is given in \href{https://tel.archives-ouvertes.fr/tel-03542389/}{https://tel.archives-ouvertes.fr/tel-03542389/}}. \\

\begin{algorithm}[t]
	\footnotesize
	\caption{Function \textit{isSeed}}
	\label{alg:isSeed}
	\SetKwInOut{Input}{input}%
	\SetKwInOut{Output}{output }%
    \Input{
        $P$, a process tree ;\\
        \ $P'$, a process tree.
    }
    \Output{
    \ $res$, a boolean.}
	\BlankLine
	 \eIf{$depth(P') \leq depth(P)$}{
	    \If{$P$ is a leaf}{
	         $res \leftarrow (P' = P)$\;
    	 }
         \ElseIf{$P = x(P_1, P_2)$ with $P_1$, $P_2$, two subtrees}{
            \If{$x$ is a constraining operator \emph{AND} ($\mathit{isSeed(P', P_1)}$ \emph{OR} $\mathit{isSeed(P', P_2)}$)}{
	         $res \leftarrow true$\;
	        }
             \ElseIf{$P' = x(P'_1, P'_2)$ where $P'_1$, $P'_2$ two subtrees \emph{AND} $\mathit{isSeed(P'_1, P_1)}$ \emph{AND} $\mathit{isSeed(P'_2, P_2)}$
            }{
             $res \leftarrow true$\;
             }
         }
	 }
     {$res \leftarrow false$\;}
\end{algorithm}

\subsubsection{Loop Seeds}
The loop operator exhibits the particular property that its behavior cannot appear without involving that of another operator: the sequence. Indeed, when respecting the compactness property (see section \autoref{sec:pattern_discovery}), the behavior of $loop(a,b)$ is $\Sigma(loop(a,b)) = \{\langle a, b, a \rangle, \dots\}$. If a trace contains this behavior, it automatically contains $\Sigma(seq(a,b)) = \{\langle a, b \rangle\}$. Therefore, outputting both $seq(a,b)$ and $loop(a,b)$ is one of the redundancies we strive to avoid. 

If two patterns P and P' are identical, except at the level of a node, which is a sequence in P', but a loop in P, we call $P'$ a loop seed.
If the output set contains two patterns P and P', where P' is a loop seed of P, then we remove $P'$, as the sequence operation in $P'$ is already accounted for in  $P$ through the loop.

\subsubsection{Minimality of the resulting set}
The set of discovered frequent patterns is minimal if it does not contain a pair (P, P') such that P' is a seed of P.

In the post-processing step, we ensure that the output set of patterns is minimal. The regular seeds of a pattern are eliminated by the discovery algorithm. 
In order to discard the alternative seeds,  we compare the patterns pairwise. For two process trees, $P'$ the less deep one and $P$, we test if $P'$ belongs to $f(P)$. This is realized through the function \textit{isSeed(P', P)} in $\autoref{alg:isSeed}$. We remove it if that is the case.  

Detecting a loop seed is performed syntactically. If considering two process trees, $P'$ and $P$, the only difference between their representative words is a loop operator in $P$ instead of a sequence operator as in $P'$ then, $P'$ is a loop seed. The representative word is constructed by pre-order traversal of the nodes, outputting activities and operators.

\subsection{Trees Equivalency}
\label{sec:equiv}
Not only do we eliminate non maximal trees in the post-processing operation but we also detect equivalent trees. We distinguish two notions of equivalency between behavioral patterns:

\begin{itemize}
\item\textbf{Syntactical equivalency:} This equivalence appears as a direct result to the existence of symmetrical operators in the process trees. Indeed, the same order-insensitive process tree can be depicted in more than one way considering the interchangeability of the children of every symmetrical operator.
\item\textbf{Behavioral equivalency:} two behavioral patterns are behaviorally equivalent if their languages are equal. In other words, the set of traces they generate are the same.
\end{itemize}

In order to achieve the objective behind our post-processing, we eliminate redundant equivalent trees. It is to be noted that syntactically equivalent process trees were already taken care of in the main algorithm of COBPAM. This was handled through the definition of the unified representative word mentioned earlier (constructed by pre-order traversal of the nodes of the tree) that is unique to all syntactically equivalent trees.

Post-processing, in turn, makes sure to avoid outputting trees exhibiting a behavioral equivalency. Accordingly, all trees outputted are behaviorally unique. Moreover, if a redundant sub-pattern  is removed, all its equivalent trees are removed too. Behaviorally equivalent trees are detected by comparing their languages.

\subsection{Visualization}
\label{sec:viz}
In order to navigate the behavioral patterns uncovered by our method, we conceive a graph-based visualization. Not only does it ensure a global and simultaneous view of all patterns, it also harbors interesting relationships between them.

In the visualization graph, nodes represent patterns  obtained after the post-processing step. 
Inspired from algebra intervals and taking into account boundaries of patterns, we define two types of relationships : follows and span. 
If the relative frequency of the traces inside the log where the appearance of $P_2$ is after the appearance of $P_1$ is greater than the threshold $\tau_f$ then we say that $P_2$ follows $P_1$. For the span relationship, if the relative number of traces inside the log where the interval defined by the boundaries of $P_1$ fits inside the interval defined by the boundaries of $P_2$ is greater than the threshold $\tau_s$ then we say that $P_2$ spans $P_1$. (See \cite{DBLP:phd/hal/Acheli21} for formal definitions.) 
In an analogous way, we define inter-follows and inter-span relationship, by considering  the relative number of traces inside the shared traces between $P_1$ and $P_2$, instead of all the log.
These two types of relationships can give supplementary insights when two patterns do not frequently appear together (and thus cannot satisfy the follows and span relationship constraint on all the log), but when  they \textit{do} occur together, they appear in the same order or one spans the other one.

Based on these relations, we define a visualization graph, whose nodes are frequent patterns and edges represent relationships between patterns. The maximum number of different edges linking the same two nodes is thus four. We apply a transitive reduction on the relationships \emph{Follows} and \emph{Spans} as they are transitive relationships. An example of a visualisation graph resulting from our experiments is showed in \autoref{fig:acobVisZ}.

The resulting graph can be seen as a descriptive, hierarchical and simplified process model of $L$ where instead of activities, behavioral patterns are used as nodes. The hierarchy is induced by the \textit{Spans} and \textit{Inter-spans} relationships. Moreover, the relationships in this graph operate at two levels of granularity. One between patterns as discussed in the definitions above and the other between activities themselves inside the patterns. This offers even more analytical information. Indeed, in the case of "spaghetti" processes, such a structure is highly useful in reducing the difficulty of the analysis while fostering interesting insights.

\section{Experimental Evaluation}
\label{sec:exp}

This section presents the evaluation of the alignment growth procedure and of the post-processing step by showing their effects on the efficiency and the effectiveness of the behavioral pattern algorithm COBPAM.
 Moreover, we show how the relationships between patterns can be explored using the visualisation module. We start by presenting our setup.

\subsection{Setup and Datasets}
\label{sec:setup}

We implemented the alignment growth algorithm and the postprocessing step and included them in  a new version of the COBPAM algorithm, called ACOBPAM. They are available as as a plugin in the ProM framework~\cite{vanDongen2005TheSupport} in the same package as COBPAM (package \textit{BehavioralPatternMining}). Due to the limited capabilities of the ProM framework in terms of graph drawing, the visualization module has been implemented as a separate Python script fed with a JSON model of the visualization graph (containing behavioral patterns and their interdependencies). The Python script is publicly available\footnote{\url{https://github.com/Alchimehd/ACOBPAM-Vis-}}.
Note that we ran the experimental evaluation on a PC with an i7-2.2Ghz processor, 16GB RAM and Windows 10.

Our experiments used the following real-world event logs. They cover different domains and are publicly available.\footnote{\url{https://data.4tu.nl/search?q=:keyword:"real\%20life\%20event\%20logs"}} Moreover, they are related to flexible processes and are of reasonable size to be explored.

\begin{itemize}
\item Sepsis: A log of a treatment process for Sepsis cases in a hospital. It contains 1050 traces with 15214 events that have been recorded for 16 activities.
\item Traffic Fines (T F): A log of an information system managing road traffic
fines, containing 150370 traces, 561470 events, and 11 activities.
\item WABO: A log of a building permit application process in the Netherlands. It contains 1434 traces with 8577 events, recorded for 27 activities.
\item BPI\_2019S1 (B\_S1): A 30\% sample of the BPI Challenge 2019
log.
The log belongs to a multinational company working in the area of coatings and
paints and records the purchase order handling process. The sample regroups 479845
events distributed over 75519 traces with 41 event classes.
\item BPI\_2019S2 (B\_S2): A 40\% sample of the previously mentioned event log, BPI Challenge 2019. The sample regroups 670583
events distributed over 105962 traces with 42 event classes.

\end{itemize}

The discovery algorithms were configured with a threshold of 0.7 for the support and precision and
two for the maximal depth of the patterns, unless stated otherwise.

Concerning the visualization graph, we chose to set the relationships support thresholds to 0.7.

\subsection{Efficiency}

Our aim is to evaluate the alignment growth method and assess how much it shortens runtimes. 

We compared the ACOBPAM algorithm including the alignment growth method with the version  using the classical alignment method (COBPAM). Note that the post-processing and visualization stepts of ACOBPAM are not inlcuded, as they are orthogonal and depend only on the returned set. 

The comparison results are presented in \autoref{tab:execTimesAd}. For each algorithm and each log, there were two executions with a depth of two for the discovered trees at first and then with a depth of three. The gain in execution time with respect to the initial runtime is also given. For Traffic Fines, there was no reduction in runtimes since all the trees evaluated contained the \textit{xor} operator which is treated classically. Yet, for the most time consuming logs, there was a highly substantial decrease, between 60\% and 66\%, in execution times, proving the efficiency of the alignment growth algorithm. However, the extraction of behavioral patterns for Sepsis under depth 3 was not possible.

Furthermore, for the most demanding logs, the gap in runtimes between COBPAM and ACOBPAM widens when we increase the depth. That is only logical because the new trees constructed of depth 3 are way more numerous than those of depth 2. The entire structure of each tree needs to be evaluated on each complete trace in COBPAM. There is an exponentiality property on two aspects when moving from depth two to three in COBPAM: the number of trees to evaluate increases exponentially, but also the runtime of the alignment operation on the new trees (of depth three) increases exponentially. That is because the classical alignment algorithm is hard with respect to the complexity/size of the tree. On the contrary, the exponentiality in ACOBPAM still applies on the number of new trees to evaluate but is less strong with respect to the alignment as, mostly, the alignments are grown incrementally.

Yet, for BPI\_2019S2, the gap didn't widen but shortened on the contrary. We thoroughly investigated this behavior and we observed that this particular log consumed a lot of memory. The garbage collector which is a java program that frees unused space to increase the memory available was called a lot. As such, more time was dedicated to freeing memory than to executing the algorithm. Finally, the supposed gain in runtime thanks to the alignment growth method was spoiled by the garbage collector interventions.
\begin{table*}[h!]
\footnotesize
    \centering
    \begin{tabular}{l l r @{\hspace{2em}} r @{\hspace{2em}} r @{\hspace{2em}} r @{\hspace{2em}} r @{\hspace{2em}} r @{\hspace{2em}}}
    \toprule
    & & Sepsis  & Traffic Fines & WABO & BPI\_2019S1 & BPI\_2019S2 \\
     \midrule
    \multirow{3}{*}{Depth 2} & COBPAM & 34s & 13s & 2s & 2.45mn & 5.8mn \\
     & ACOBPAM & 12s & 13s & 1s & 59s & 2mn \\
     & Runtime decrease & 65\% & 0\% & 50\% & 60\% & 66\% \\
     \midrule
     \multirow{3}{*}{Depth 3} & COBPAM & >24h & 13s & 3s & 5.81mn & 14.9mn  \\
     & ACOBPAM & >24h & 13s & 2s & 98s & 6.9mn  \\
     & Runtime decrease & N/A & 0\% & 33\% & 72\% & 49\%\\
     \bottomrule
    \end{tabular}
    \caption{Execution times of COBPAM and ACOBPAM (for a depth parameter of two and three)}
        \label{tab:execTimesAd}
\end{table*}

Finally, we also compared ACOBPAM with LPM discovery. While the two algorithms use different notions of patterns support and frequency and cannot be fairly compared, the experiments give an indication of execution times needed to mine patterns of same size by both algorithms. For the implementation of LPM discovery in ProM, we chose the default parameters for the mimimum support and quality (determinism), as suggested by the authors (these are automatically calculated by the tool, depending on the event log characteristics, such that the number of events and activities). The size of LPMs to discover was set to 4 which corresponds to patterns with depth 2 under our algorithm. On another hand, ACOBPAM discovers all frequent patterns under the set depth. LPM Miner doesn't discover the whole set but limits its size. As such, in its parameters, we fix the size of the set to discover to the maximal value: 500 .

We give in \autoref{tab:acobLpmTimes} the decrease in runtime observed with respect to  LPM Discovery algorithm. We can see that the execution times of our algorithm are order of magnitude lower than the state of the art. Indeed, we divided the runtime of Sepsis by 100 and that of Traffic Fines by 1000.  Moreover, LPM was not able to return any results for three out of the five evaluated logs (Traffic Fines, BPI\_2019S1 and BPI\_2019S2).

\begin{table}[h]
\footnotesize
 \centering

  \begin{tabular}
{l r @{\hspace{2em}} r @{\hspace{2em}} r @{\hspace{2em}} r @{\hspace{2em}} r @{\hspace{2em}} r @{\hspace{2em}} r}

    \toprule
     & Sepsis  & T F & WABO & B\_S1 & B\_S2\\
     \midrule
      Runtime \% & 99\% & N/A & 99.9\% & N/A & N/A \\     \bottomrule
    \end{tabular}
   \caption{Runtime decrease with respect to\\ the state-of-the-at algorithm for mining behavioral patterns (LPM)}\label{tab:acobLpmTimes}
\end{table}

Next, we report in \autoref{tab:postProcTimes} the runtimes of the post-processing operation and visualization graph construction for both considered depths. The most time consuming part is the post-processing operation. Its runtime increases with respect to the initial number of trees returned.  Indeed, there is a need for pairwise comparison to check equality of languages and existence of generalized maximality seeds. This can be verified by analyzing \autoref{tab:postProcTrees}. On another hand, the computation of the language, the detection of the loop seeds and the detection of the alternative seeds are all computationally hard with respect to the depth. That is because the tree gets exponentially more complex and the number of its subtrees (which are used in the definition of alternative seeds) grows exponentially too when it gets deeper. This last statement is confirmed in the depth 3 runtime of BPI\_2019S2 seen in \autoref{tab:postProcTimes}. It is the time needed for the pairwise comparison of the 79 trees returned. However, its value: 1.8mn is almost equivalent to the time needed to compare 597 trees for Sepsis under depth 2 (1.95mn).

\begin{table}[h!]
\footnotesize
\centering

    \begin{tabular}{l r @{\hspace{2em}} r @{\hspace{2em}} r @{\hspace{2em}} r @{\hspace{2em}} r @{\hspace{2em}} r @{\hspace{2em}} r}
    \toprule
     Depth & Sepsis  & T F & WABO & B\_S1 & B\_S2\\
     \midrule
     2 & 1.95mn & 0s & 3s & 14s & 19s \\

     3 & N/A & 0s & 1s & 14s & 1.8mn \\
     \bottomrule
    \end{tabular}
 \caption{Runtimes of the post-processing operation \\ and visualization graph generation}
    \label{tab:postProcTimes}
\end{table}

\subsection{Effectiveness}
\subsubsection{Quantitative Effectiveness}

We give in \autoref{tab:postProcTrees} the number of trees returned before and after the application of post-processing for each considered depth along with the ratio between the initial number and the final one. Except for Traffic Fines which returned initially only one tree, the post-processing operation managed to reduce the number of trees returned by ACOBPAM and appearing during the visualization to between 35\% to 73\% their initial number. This demonstrates that the number of redundant trees in the first set is non negligible and that the generalized maximality with the equivalency study are indeed welcome and necessary. An other interesting observation is that the gain is higher in depth 3; that is because, each tree of such depth has a higher number of alternative/loop seeds and equivalent trees which are undoubtedly initially returned and consequently eliminated.
\begin{table}[h!]
\footnotesize
  \centering
    \begin{tabular}{l l r @{\hspace{2em}} r @{\hspace{2em}} r @{\hspace{2em}} r @{\hspace{2em}} r @{\hspace{2em}} r @{\hspace{2em}} r}
    \toprule
     Depth  & & Sepsis  & T F & WABO & B\_S1 & B\_S2 \\
     \midrule
      \multirow{2}{*}{2} & O & 597 & 1 & 32 & 36 & 37 \\
      & W & 439 & 1 & 16 & 26 & 27 \\
      & Ratio & 73\% & 100\% & 50\% & 72\% & 73\% \\
     \midrule

      \multirow{2}{*}{3} & O & N/A & 1  & 26 & 47 & 79  \\
     & W & N/A & 1  & 9 & 26 & 31  \\
     & Ratio & N/A & 100\% & 35\% & 55\% & 39\% \\
     \bottomrule
    \end{tabular}
     \caption{Number of returned trees w/o Post-processing}
      \label{tab:postProcTrees}
\end{table}

\subsubsection{Qualitative Effectiveness}

We show in \autoref{fig:cobNRes}, a subset of the trees returned by COBPAM for WABO. It corresponds to the trees returned before the post-processing operation. Arguably, since trees \autoref{tree:p1} and \autoref{tree:p3} represent the alternative seeds of \autoref{tree:p2} and \autoref{tree:p4} respectively, only \autoref{tree:p2} and \autoref{tree:p4} were kept in the final set returned and visualized in ACOBPAM.

\begin{figure}
\begin{center}
\subfloat[]{
\label{tree:p1}

    \begin{tikzpicture}[thick,scale=0.7, every node/.style={transform shape}]
        \tikzset{edge from parent/.style= {draw, edge from parent path={(\tikzparentnode.south) -- +(0,-6pt) -| (\tikzchildnode)}}}
		\tikzset{level 1/.style={level distance=40pt}}
		\tikzset{level 2/.style={level distance=40pt}}
		\tikzset{level 3+/.style={level distance=40pt}}
    	\Tree [.\node[circle, draw]{\small \textit{seq}}; [.\node[rectangle, draw] {\small{CRec}}; ] [.\node[circle, draw]{\small \textit{and}}; [.\node[rectangle, draw] {\small {T10}} ; ] [.\node[rectangle, draw] {\small {T05}} ; ] ] ];

	\end{tikzpicture}
}\quad\quad\quad\subfloat[]{
\label{tree:p2}

     \begin{tikzpicture}[thick,scale=0.7, every node/.style={transform shape}]
        \tikzset{edge from parent/.style= {draw, edge from parent path={(\tikzparentnode.south) -- +(0,-6pt) -| (\tikzchildnode)}}}
		\tikzset{level 1/.style={level distance=40pt}}
		\tikzset{level 2/.style={level distance=40pt}}
		\tikzset{level 3+/.style={level distance=40pt}}
    	\Tree [.\node[circle, draw]{\small \textit{seq}}; [.\node[circle, draw] 
     {\small \textit{seq}};
     [.\node[rectangle, draw]{\small{CRec}}; ] [.\node[rectangle, draw]{\small {T02}}; ] ] [.\node[circle, draw]{\small \textit{and}}; [.\node[rectangle, draw]{\small {T10}}; ] [.\node[rectangle, draw]{\small {T05}}; ] ] ];
	\end{tikzpicture}
}
\quad\quad\quad\subfloat[]{
\label{tree:p3}

 \begin{tikzpicture}[thick,scale=0.7, every node/.style={transform shape}]
        \tikzset{edge from parent/.style= {draw, edge from parent path={(\tikzparentnode.south) -- +(0,-6pt) -| (\tikzchildnode)}}}
		\tikzset{level 1/.style={level distance=40pt}}
		\tikzset{level 2/.style={level distance=40pt}}
		\tikzset{level 3+/.style={level distance=40pt}}
    	\Tree [.\node[circle, draw]{\small \textit{seq}}; [.\node[rectangle, draw]{\small{CRec}}; ] [.\node[circle, draw]{\small \textit{seq}}; [.\node[rectangle, draw]{\small {T06}}; ] [.\node[rectangle, draw]{\small {T10}}; ] ] ];

	\end{tikzpicture}
}\quad\quad\quad\subfloat[]{
\label{tree:p4}

 \begin{tikzpicture}[thick,scale=0.7, every node/.style={transform shape}]
        \tikzset{edge from parent/.style= {draw, edge from parent path={(\tikzparentnode.south) -- +(0,-6pt) -| (\tikzchildnode)}}}
		\tikzset{level 1/.style={level distance=40pt}}
		\tikzset{level 2/.style={level distance=40pt}}
		\tikzset{level 3+/.style={level distance=40pt}}
    	\Tree [.\node[circle, draw]{\small \textit{seq}}; [.\node[circle, draw]{\small \textit{seq}}; [.\node[rectangle, draw]{\small{CRec}};] [.\node[rectangle, draw]{\small {T02}}; ] ] [.\node[circle, draw]{\small \textit{and}}; [.\node[rectangle, draw]{\small {T06}}; ] [.\node[rectangle, draw]{\small {T10}}; ] ] ];
	\end{tikzpicture}
}
\end{center}
\caption{Behavioral Patterns mined by COBPAM for WABO (CRec: Confirmation of Receipt)}
\label{fig:cobNRes}
\end{figure}

\normalsize

\subsection{Visualisation}

\begin{figure}[t]
\centering
\includegraphics[width=0.4\columnwidth]{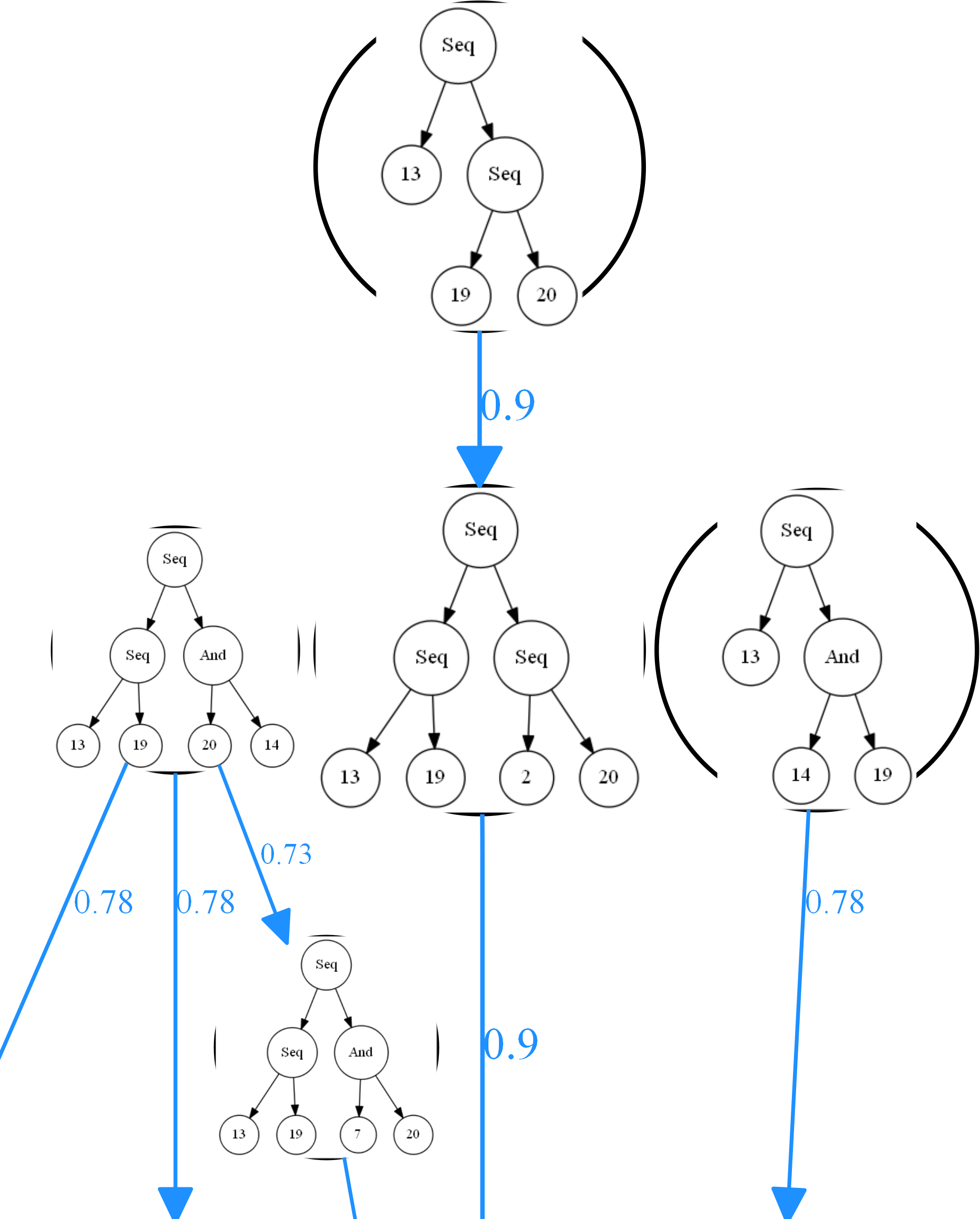}
\caption{Visualisation graph of the patterns returned by WABO (support threshold of 0.7, depth of 2)[Zoomed-in view]}
\label{fig:acobVisZ}
\end{figure}

We give in \autoref{fig:acobVisZ} a zoomed-in version of  the generated visualization graph for WABO (support threshold of 0.7 and depth of 2). 

The behavioral patterns are represented in their tree form inside the nodes of the graph. The node is circular and its size is proportional to the support of the pattern. The support can be displayed by hovering over it. There are four edge shapes and colors to depict each of the four relationships: \textit{Follows, Inter-follows, Spans and Inter-spans relationships}. Obviously, this graph contains only \textit{Spans relationships}.
Note that the edges are weighted with the supports of the relationships.

Moreover, the activities inside the trees are replaced by numbers to avoid cluttering and complexity which is the opposite of what we expect from the visualization graph.

\section{Conclusion}
\label{sec:conclu}

In this paper we improve state-of-the-art techniques for behavioral pattern mining by addressing two of their limitations: the inefficiency of the  evaluation of pattern candidates due to costly alignments and the overwhelming number of outputted patterns.  We achieve efficient mining by integrating an incremental evaluation of pattern candidates and we improve the patterns’ utility by effective means for filtering and visualization. The filtering avoids outputing redundant information, while the visualization framework  offers an interactive view on the relations between behavioral patterns.

We evaluated our techniques with diverse real-world
datasets. Our results indicate that our algorithm for incremental
evaluation can speed up discovery by up to a factor of
3.5. Also, our post-processing procedure reduces the number
of discovered patterns to  35\%-75\% their initial size depending on the dataset.

In future work, to further facilitate the exploration of the discovered patterns, ranking features could be introduced using preferences and utility measures. On another hand, while the current implementation of the algorithm is able to extract small patterns (3-4 activities) in reasonable time, in order to be able to handle huge volume of logs and output maximal patterns of any size, we will investigate using parallel computing frameworks.
Finally, since behavioral patterns can be seen as a summary of the underlying process, they could
be used as features in various mining and machine learning tasks, like trace clustering and predictive monitoring.

\bibliographystyle{elsarticle-num}
\bibliography{references}







\end{document}

%% file: introduction.tex
\section{Introduction}

Process Mining has become an integral asset for organizations
to deal with their processes~\cite{VanderAalst2016ProcessAction}.
By combining approaches from business process management with data science
techniques, manifold opportunities emerge to gain
insights for business processes using event logs recorded during their
execution. Such logs are leveraged to discover process
models, to check conformance with reference models, to enhance
models with performance information, and to provide operational
decision support.

Since process models denote the starting point for a wide range of process
improvement initiatives~\cite{DBLP:books/sp/DumasRMR18}, process
discovery algorithms received particular attention in the
field~\cite{DBLP:journals/tkde/AugustoCDRMMMS19}. Traditional
algorithms~\cite{vanderAalst2004WorkflowLogs,
Burattin2012HeuristicsData,Leemans2013LNCSApproach,
DBLP:conf/icdm/AugustoCDR17,DBLP:journals/dss/BrouckeW17} aim at the
construction of a \emph{single} model for the \emph{whole}
behavior of a process. While these algorithms cope well with relatively
structured processes, they fail to provide insights for scenarios with large
variability between process executions. In such cases, it is more meaningful to
construct \emph{multiple} models, each capturing some \emph{partial} behavior,
which jointly describe the process. To operationalize this idea, one may
split the event log with methods for trace
clustering~\cite{DBLP:conf/bpm/BoseA10,
	DBLP:conf/icsoc/SunBW17,DBLP:journals/kais/KoninckNBBSW21}
 before applying a
traditional discovery algorithm, or construct a set of models that capture
behavioral regularities observed for all traces, captured either as declarative
constraints~\cite{Maggi2011User-guidedModels,
DiCiccio2016} or behavioral
patterns~\cite{Tax2016MiningModels,acheli2021discovering}.

Approaches to discover behavioral regularities observed for all traces have the
advantage of, compared to trace clustering, enabling a more fine-granular
exploration of a process. Even regularities observed only for
small parts of traces,
which would \emph{not} yield clusters of complete traces, can be
captured by behavioral patterns to support understanding and analysis of the
underlying process. At the same time, the representation of regularities
through behavioral patterns is closer to well-established procedural languages
for process modeling than a formalization as declarative constraints.

Behavioral patterns are often represented as process trees, i.e., tree
structures that comprise control-flow operators as non-leaf nodes and the
process' activities as leaf nodes. As such, process trees go beyond sequential
patterns and support complex behavior, including repetition,
choices, and concurrency.
Consider the
example in \autoref{fig:runExample}, where \autoref{fig:log} shows an
event log of a treatment process of a patient in a hospital. Each row, called
trace, relates to one execution of the process. For this log, the pattern in
\autoref{fig:tree}, is frequent, i.e., it is present in 9 out of 12 traces. It
highlights that blood tests (BT) are typically followed by both, the checkout
(CO) and the retrieval of belongings (RB), in either order.

\begin{figure}[t]
	\centering
	\subfloat[]{\label{fig:log}
		\scriptsize
		\begin{tabular}{l @{\quad} l}
			\toprule
			Trace ID & Event Sequence \\
			\midrule
			1& EI ET PS ED \textcolor{red}{BT} BT GP TD SW
			\textcolor{red}{CO} \textcolor{red}{RB} \\
			2& ET EI CV XS \textcolor{red}{BT}
			SW CS D
			\textcolor{red}{RB} \textcolor{red}{CO}\\
			3& CI PS CV I  \textcolor{red}{BT} XS
			SW E
			\textcolor{red}{CO} \textcolor{red}{RB} I \\
			4& CI CV PS XS \textcolor{red}{BT} D
			SW CS GP
			\textcolor{red}{RB} \textcolor{red}{CO}\\
			5& CI PS EI ED I XS GP TD CV \\
			6& EI ET PS ED \textcolor{red}{BT} GP TD
			\textcolor{red}{CO} \textcolor{red}{RB}\\
			7& ET EI CV XS \textcolor{red}{BT} CS D
			\textcolor{red}{CO} \textcolor{red}{RB}\\
			8& CI PS CV I  \textcolor{red}{BT} XS E
			\textcolor{red}{CO} \textcolor{red}{RB} I \\
			9& CI CV PS XS \textcolor{red}{BT} D CS GP
			\textcolor{red}{CO} \textcolor{red}{RB}\\
			10& CI PS EI ED I XS GP TD CV \\
			11& ET PS ED \textcolor{red}{BT} GP TD
			SW
			\textcolor{red}{CO} \textcolor{red}{RB} \\
			12& CI PS EI ED I XS GP TD CV \\
			\bottomrule
			\addlinespace[.2em]
			\multicolumn{2}{l}{
				\tiny{
				SW: Meet with social worker,
				CO: Checkout,
				BT: Blood test,
				CI: Check-In,
				GP: Give prescription,
			}}\\
			\multicolumn{2}{l}{
				\tiny{
				CS: Recheck sec. number,
				EI: Emergency intubation,
				ET: Emergency transfusion,
			}}\\
\multicolumn{2}{l}{
\tiny{
				PS: Process sec. number,
				CV: Check Vitals,
				XS: X-ray Scan,
				TD: Temp. Diagnosis,
				D: Diagnosis,
			}}\\
\multicolumn{2}{l}{
\tiny{
				ED: Emergency defibril.,
				I: Infusion,
				E: Echography,
				RB: Retrieve belongings
			}}
		\end{tabular}
	}\quad
	\subfloat[]{\label{fig:tree}\centering
		\begin{tabular}{c}
			\begin{tikzpicture}
			\tikzset{level 1/.style={level distance=24pt}}
			\tikzset{level 2/.style={level distance=20pt}}
			\tikzset{level 3+/.style={level distance=20pt}}
			\Tree [.\textit{seq} [.BT ] [.\textit{and} [.CO ] [.RB ] ] ]
			\end{tikzpicture}
		\end{tabular}
	}

	\vspace{-.5em}
	\caption{(a) Event log;
		(b) behavioral pattern; The pattern has a
	    support > 0.7 (i.e., occur in more than eight out of 12 traces).
    }
	\label{fig:runExample}
	\vspace{-1em}
\end{figure}

Several algorithms for discovering behavioral patterns from event logs have
been proposed in the
literature~\cite{Tax2016MiningModels,acheli2019efficient, acheli2021discovering,Peeva21}.
However, despite
notable
advancements in recent years, existing approaches still show some general
inefficiency and ineffectiveness. The former is due to two reasons. First,
the set of pattern candidates grows exponentially. Second, the evaluation
of the quality of each pattern candidate is based on
alignments~\cite{Adriansyah2014AligningBehavior} between the process tree
and the traces, which induces an exponential runtime
complexity~\cite{DBLP:books/sp/CarmonaDSW18}. However, current algorithms
address solely the first reason of
inefficiency~\cite{acheli2021discovering}, \cite{Peeva21}.
Also, effective use of
the mining results is hindered by the sheer number of the patterns obtained in
practice. Independent of goal functions to guide the mining
procedure~\cite{DBLP:journals/is/TaxDSAN18},
current notions of pattern maximality~\cite{acheli2019efficient} provide a
relatively strict characterization of behavioral redundancy and, hence, fail to
reduce the number of patterns considerably.

In this paper, we address the above issues.
We achieve efficient
mining by integrating an incremental evaluation of pattern
candidates in the COBPAM algorithm~\cite{acheli2019efficient}, a
state-of-the-art technique for pattern mining. In addition, we improve the
patterns' utility by effective means for filtering
and visualization. We summarize our contributions, coined Advanced COBPAM, as follows:

\begin{compactenum}[(1)]
\item We devise an incremental algorithm for the evaluation of pattern
candidates in the mining procedure. It assesses patterns by
``growing'' alignments computed earlier for smaller patterns, thereby saving
computational effort.

\item We show how to reduce the number of obtained patterns through
post-processing. Beyond pattern equivalence, this step exploits a novel
notion of pattern maximality to capture if one pattern can be inferred from
another one.

\item To further support the exploration of behavioral patterns, we present a
visualization framework. It offers an interactive view on the relations between
behavioral patterns.

\end{compactenum}
We evaluated our techniques with diverse real-world datasets. Our results
indicate that our algorithm for incremental evaluation can speed up discovery
by up to a factor of 3.5. Also, our post-processing procedure reduces
the number of discovered patterns to
35\%-75\%, depending on the dataset.

In the remainder, \autoref{sec:relatWorks}
reviews related work, while \autoref{sec:cobpam} provides the
necessary background for our work.

\autoref{sec:aligrow} presents our algorithm for incremental evaluation of
pattern candidates. Then, in \autoref{sec:postpro}, we present a
post-processing procedure to reduce the number of returned patterns. A
framework for interactive exploration of patterns is introduced in
\autoref{sec:viz}. An experimental
evaluation is presented in \autoref{sec:exp}, before we conclude in
\autoref{sec:conclu}.

%% file: related_work.tex
\section{Related Work}
\label{sec:relatWorks}

The discovery of frequent behavioral patterns in
an event log connects several research areas,
including sequential pattern mining and
process discovery. In this section, we review related algorithms from both areas along with a short discussion on alignments.

\medskip
\noindent
\textbf{Sequential pattern mining.} Given a database of sequences of data
elements, sequential pattern mining aims at the extraction of frequently
recurring (sub-)sequences of elements~\cite{Fournier-Viger2017AMining}. While
there have been decades of research
on pattern mining algorithms, most of them proceed incrementally. For instance,
the GSP algorithm~\cite{Srikant1996MiningImprovements}
combines pairs of sequential patterns of length $k$ to obtain
patterns of length $k+1$.

Moreover, various approaches to optimize the runtime behavior of sequential
pattern mining have been proposed. For instance, the PrefixSPAN
algorithm~\cite{Pei2004MiningApproach} presents the notion of a projected
database to evaluate the pattern candidates on the minimal number of rows
possible. In fact, the projected database is not only a subset of the table but
also a shifted copy of the rows, i.e., a subset of tails of the
tuples. Orthogonally, measures such as maximality have been proposed to guide
the search for the most relevant patterns~\cite{Fournier-Viger2017AMining}.

Optimizations that aim at an efficient discovery and measures to characterize
relevance may be lifted to the setting of behavioral patterns. For
instance, the idea of combining pairs of patterns of length $k$ to obtain
patterns of length $k+1$ has been adopted for behavioral
patterns~\cite{acheli2019efficient}.

\medskip
\noindent
\textbf{Process discovery.}
Algorithms for process discovery construct models with rich semantics that
feature not only sequential dependencies, but concurrency, exclusive choices,
or repetitive behavior from event logs, i.e., collections of
traces~\cite{DBLP:journals/tkde/AugustoCDRMMMS19}. While most approaches focus
on the
discovery of a single end-to-end model, their
application in scenarios with relatively unstructured behavior will not yield
useful insights~\cite{VanderAalst2016ProcessAction}.
To tackle this problem, behavioral pattern mining has been introduced. The
notion of behavioral patterns was first proposed
in~\cite{Tax2016MiningModels} under the term Local Process Models
(LPMs). Since then, various proposals have been presented to increase the
efficiency of pattern discovery, given the exponential runtime of common
discovery algorithms. This primarily includes pruning rules as well as
particular strategies to explore the space of pattern
candidates~\cite{acheli2019efficient,Peeva21}. For instance, while LPM Mining includes pruning, it does so by camparing a candidate to one tree while COBPAM compares to a pair of them. This, in turn, allows to multiply the pruning opportunities. Nonetheless, the runtime complexity
induced by the evaluation of the quality of each pattern candidate was not yet
subject to any optimization.

Moreover, while behavioral patterns proved to be useful in many application
domains, such as healthcare~\cite{DBLP:conf/icpm/PijnenborgVFLG21}, their
effective use is often hindered by the sheer number of discovered patterns. As
such, it was suggested to employ notions of compactness and maximality to avoid
the discovery of redundant patterns~\cite{acheli2019efficient}, to guide the search by measures that capture the interestingness of a
pattern~\cite{DBLP:journals/is/TaxDSAN18} or implying the user in an interactive, iterative and multi-dimensional selection of patterns~\cite{InteractiveLPM}.

These existing notions and measures, however, are based on a relatively strict
characterization of behavioral redundancy. In practice, therefore, automatic discovery
results still include a large number of patterns, many of which provide
redundant insights, while interactive discovery requires user intervention.

A pruning approach inspired from compressed sequential pattern mining \cite{TaxThesis} selects patterns (local process models) whose combination into a global process model (PN) achieves a trade-off between non-redundancy and coverage of events in the log. Our approach discovers the minimal set of maximal behavior patterns of a given size k and  shows temporal relationships between them.

\medskip
\noindent
\textbf{Alignments}
Alignments are a tool used in conformance checking to elicitate and uncover discrepancies between an established model and an event log. The A* algorithm and different cost functions~\cite{de2013aligning, koorneef2017automatic, burattin2017framework} are used to compute the optimal alignment, that is, the closest fitting between a process and a trace. However, the main issue with alignments is their exponential complexity. Consequently, in order to cater for large scale event logs and/or models, some approaches introduce a divide-and-conquer strategy where they aim to align parts of the model instead of the whole one~\cite{munoz2014single, song2016efficient, verbeek2014decomposed}. 
Particularly, an alignment repairing procedure is presented in ~\cite{van2020repairing} which uses previously aligned models in order to align newer models that are similar. ACOBPAM uses such a divide-and-conquer strategy and computes alignements based on similar, smaller, models too. However, in the latter approaches, the techniques do not guarantee the optimality of the alignments returned while ACOBPAM discovers an optimal alignment that extracts the exact behavioral pattern embedded in the trace.

%% file: background.tex
\section{Background on Behavioral Patterns}
\label{sec:cobpam}

We discuss event logs and {process trees to model behavioral patterns}
(\autoref{sec:logs_trees}), before turning to the related discovery problem
and the existing COBPAM algorithm (\autoref{sec:pattern_discovery}).

\subsection{Event Logs and Process Trees}
\label{sec:logs_trees}

We start by introducing event logs.
Let $A$ be a set of \emph{activities}, and $A^*$
the set of all sequences over $A$.
As usual,
for a sequence $\sigma_1=\langle a_1,\ldots,a_n \rangle\in A^*$, we write
$|\sigma_1|=n$ for its length,
$\sigma_1(i)=a_i$, $1\leq i\leq n$, for the i-th activity, and
$\sigma_1.\sigma_2=\langle a_1,\ldots,a_n,b_1,\ldots,b_m \rangle$
for the concatenation with a sequence
$\sigma_2=\langle b_1,\ldots,b_m \rangle\in A^*$.
An interleaving of $\sigma_1$ and $\sigma_2$ is a sequence
$\sigma_3$ of length $n+m$ that is induced by a bijection $\alpha:
\{1,\ldots,{n+m}\}\rightarrow \{(s,j)\mid 1\leq s\leq 2\land 1\leq j \leq
|\sigma_s|\}$, which maps activities of $\sigma_3$ to those in $\sigma_1$ and
$\sigma_2$, while preserving their original order in $\sigma_1$ and $\sigma_2$,
i.e., for all
$1\leq i_1 < i_2 \leq n+m$ with $\alpha(i_1)=(s_1,k_1)$ and
$\alpha(i_2)=(s_2,k_2)$, it holds that $\sigma_{3}(i_1)=\sigma_{s_1}(k_1)$,
$\sigma_{3}(i_2)=\sigma_{s_2}(k_2)$, and $s_1=s_2$ implies that $k_1<k_2$.
The set of all interleavings of $\sigma_{1}$ and $\sigma_{2}$ is
denoted by $\{\sigma_1,\sigma_2\}_{\simeq}$. Additionally, a sequence $\gamma$ is a subsequence of $\sigma_1$ if and only if $\gamma$ is
a projection of $\sigma_1$, i.e., the (order preserving) removal of activities of
$\sigma_1$ yields $\gamma$, which we denote as $\gamma \sqsubseteq \sigma_1$.
Finally, $\mathit{tail}^k(\sigma) = \langle a_{min(k, n)}, a_{min(k, n) +
1},\dots, a_n
\rangle$ is the suffix or tail of $\sigma$ starting from $\sigma(k)$,
inclusive.

A \emph{trace} is a pair $(id, \sigma)$
where $\sigma \in A^*$ is a finite
sequence of activities and $id\in I$ is a trace identifier, with $I$ being the
set of possible identifiers.
Then, an \emph{event log} $L$ is a set of traces, each having a unique identifier,
	i.e., $L \subseteq I \times A^*$ where $|\{id \mid (id, \sigma) \in L\}| =
	|L|$.

Given an event log, behavioral patterns may be discovered. We represent them as
process trees, a hierarchical model defining semantics
over a set of activities. We recall the definition of a
process tree based on~\cite{Leemans2013LNCSApproach,Buijs2012ATrees}:
\begin{definition}
	A \emph{process tree} is an ordered tree, where the leaf nodes
	represent activities and non-leaf nodes represent operators. Considering a
	set
	of activities $A$, a set of binary operators $\Omega = \{seq, and, loop,
	xor\}$, a process tree is recursively defined as:
	\begin{itemize}
		\item $a \in A$ is a process tree.
		\item considering an operator $x \in \Omega$ and two process trees
		$P_1$,
		$P_2$, $x(P_1, P_2)$ is a process tree having $x$ as root, $P_1$ as
		left
		child, and $P_2$ as right child.
	\end{itemize}
\end{definition}
The above definition for process trees is slightly more restrictive than the
one presented in~\cite{Leemans2013LNCSApproach,Buijs2012ATrees}, which is
explained as follows: First, we limit ourselves to binary instead of n-ary
operators to simplify the definition of pattern discovery. However, we note
that this does not limit the expressiveness of the model, since any n-ary
operator can be represented by nesting binary operators of the respective type.
Second, we require leaf nodes to be activities and neglect a dedicated symbol
for silent steps (also known as $\tau$ steps). The reason being that the
purpose of leaf nodes that are silent steps is to capture skipping of
activities as part of an $xor$ or $loop$ operator. However, our interpretation
of a pattern is always based on a notion of a subsequence that refers to the
\emph{projection} of a sequence, i.e., relation $\sqsubseteq$ as defined above.
As such, there is no need to consider silent steps in the definition of a
behavioral pattern.

Moreover, the depth of a node (activity or operator) in the tree is the
length of the path to its root. The depth of the tree is the maximal depth
observed.

The language $\Sigma(P)$ of a process tree $P$ is a set of words, also defined
recursively.
For an atomic process tree $a \in
A$, the language is $\Sigma(a)=\{\langle a \rangle\}$. For a process tree
$x(P_1, P_2)$, the language is obtained by a function $f_x$ that
merges words of $\Sigma_1 = \Sigma(P_1)$ and $\Sigma_2 = \Sigma(P_2)$ depending
on

the operator $x$. For the sequence and concurrency operators, the function
concatenates and interleaves two words of either languages:
\begin{equation*}
f_{seq}(\Sigma_1,\Sigma_2)=\{w_1.w_2
\mid w_1\in  \Sigma_1 \land w_2\in \Sigma_2 \},
\end{equation*}
\begin{equation*}
f_{and}(\Sigma_1,\Sigma_2)= \{w \mid
w\in \{w_1,w_2 \}_{\simeq} \land w_1\in  \Sigma_1 \land
w_2\in \Sigma_2\}.
\end{equation*}
For the exclusiveness operator, the languages are unified, while
the language of the loop operator is obtained by alternating words of either
languages:
\begin{equation*}
f_{xor}(\Sigma_1,\Sigma_2)=\Sigma_1 \cup \Sigma_2,
\end{equation*}
\begin{equation*}
f_{loop}(\Sigma_1,\Sigma_2)=\{w_1.w_1'.w_2.\ldots .
w_{n-1}'.	w_n \mid n\in \mathbb{N^*}, \forall \ 1\leq i \leq n: w_i\in
\Sigma_1 \land w_i'\in  \Sigma_2\}.
\end{equation*}
For example, in \autoref{fig:runExample}, the process tree in
\autoref{fig:tree} defines the language $\{\langle BT, CO,
RB\rangle,\langle BT, RB, CO \rangle \}$.

\subsection{Discovery of Behavioral Patterns}
\label{sec:pattern_discovery}

Given an event log, only behavioral patterns that are frequent shall be
discovered. Below, we first define a notion of support to characterize frequent
patterns, before turning to the COBPAM algorithm to discover them. We also
highlight the properties in terms of pattern compactness and maximality that
are ensured by COBPAM.

\smallskip
\noindent
\textbf{Frequent patterns.}
Following the reasoning presented in~\cite{Tax2016MiningModels}, we strive for
patterns that are based on behavioral containment. For a trace $(id,\sigma)$ of
a log $L$, the behavior of a process tree $P$ is exhibited by the trace, if
there exists a word $w \in \Sigma(P)$ of the language of $P$, such that $w
\sqsubseteq
\sigma$. We refer to this word by $w = \upsilon(\sigma, P)$. Additionally, a boolean variable $\epsilon(\sigma, P)$ is associated to the containment.

Continuing with example in \autoref{fig:runExample}, we note that Trace~1 of
the event log in
\autoref{fig:log} exhibits the behavior of the process tree in
\autoref{fig:tree}. That is, the word of the language of the process tree
$\langle BT, CO,
RB\rangle$ is one of its (projected) subsequences. Note that the second
occurrence of $BT$
is not part of the projection, as the
language of the process tree does not define a repetition. Trace~8
is a counter-example, as neither of the two words of the language of the
process tree can be
derived as a subsequence.

The importance of patterns is commonly captured in terms of their support, as
follows:

\begin{definition}
\label{def:support}
Given an event log $L$, the \emph{count} of a process tree $P$ is the
number of traces that exhibit its behavior:
\begin{equation*}
\mathit{count}(P,L) = \left|\{(id, \sigma) \in L \mid \exists\ w \in
\Sigma(P): w \sqsubseteq \sigma\}\right|.
\end{equation*}
Its \emph{support} is the count over the size of the log:
\begin{equation*}
support(P,L) = \frac{\mathit{count}(P,L)}{|L|}.
\end{equation*}
\end{definition}

\smallskip
\noindent
\textbf{COBPAM algorithm.}
We build upon the COBPAM algorithm~\cite{acheli2019efficient}, a
state-of-the-art technique to discover frequent behavioral patterns in a log.
COBPAM relies on a generate-and-test approach, in which process
trees are constructed incrementally. At the core of the algorithm is the notion
of a combination. It is based on two trees with $k$ leaves that differ in one leaf. This leaf must be a potential combination leaf as defined next. 

\begin{definition}\label{Comb_leaf}
Given a process tree $P$ of depth $i$, a leaf node $a$ of depth $d$ is called potential combination leaf if $d \geq i - 1$ and there is no leaf $b$ of depth $d'$ on the right of $a$ such that $d' > d$.
\end{definition}

\begin{figure}
	\centering
	\includegraphics[scale = 0.26]{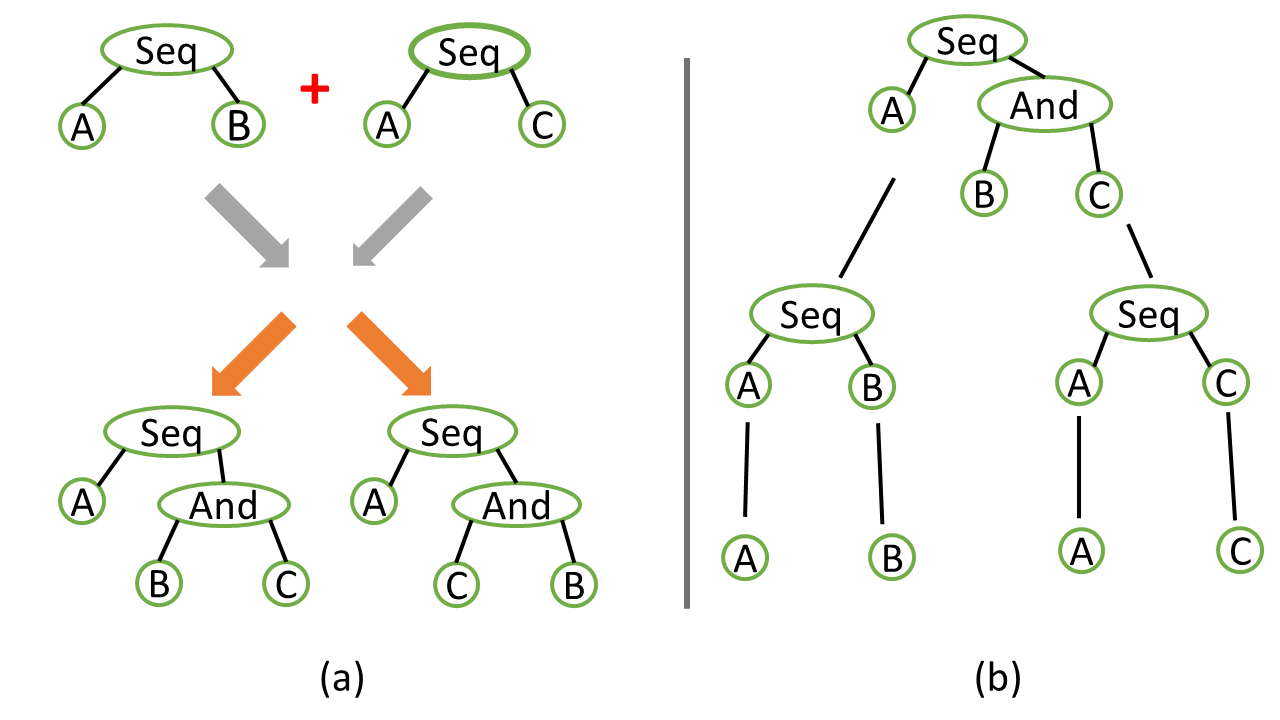}
	\caption{(a) A combination operation of two process trees with a concurrence combination operator; (b) the
		construction tree of the process tree \textit{seq(a, and(b,c))}}
	\label{fig:combinations}
	\vspace{-1.5em}
\end{figure}

We refer to these trees as regular seeds. Taking two seeds and a chosen combination operator, a combination generates
trees with $k+1$ leaves, as illustrated in
\autoref{fig:combinations}(a). When the combination operator is a sequence, concurrence or loop, this construction is monotonic in terms of the
support: A combination result can only be frequent, if the
seeds are frequent too. We call these constraining operators.

The construction realized in COBPAM exploits a partial order over the trees defined by the combinations (an inclusion order where the results of the combination operation include the
operands). In fact, for each tree $P$, there are only two regular seeds, the combination of which, yields $P$\footnote{The proof is given in \href{https://tel.archives-ouvertes.fr/tel-03542389/}{https://tel.archives-ouvertes.fr/tel-03542389/}}. Based thereon, the construction of any tree can be described in a directed acyclic graph, called
construction tree. For instance, \autoref{fig:combinations}(b) illustrates the
construction of a tree \textit{seq(a, and(b,c))} from trees comprising only a
single node through the intermediate step of trees with two leaves. The set of all construction trees form an acyclic directed graph called construction graph, which is browsed by the algorithm during the generation of the trees.

To extract patterns that are particularly useful for behavioral analysis,
COBPAM directly enables a restriction of the result to compact and maximal
patterns, defined as follows:

\textit{Compactness:} Given an event log $L$, a
process tree $P$ is \emph{compact}, if it satisfies all of the following
conditions (the last one was not considered
in~\cite{acheli2019efficient}, but follows the same spirit):
\begin{compactitem}
\item $P$ does not exhibit the choice operator as a root node. If so, the
process tree would be the union of separate trees, so that any support
threshold may be passed through the aggregation of unrelated behavior.
\item $P$ does not include a choice operator,
where, given $L$, just one child is frequent. Adding further behavior to a
frequent tree adds complexity, while the added
behavior may not even appear in the log.

\item $P$ does not contain a loop operator, such that only the behavior of the
first child appears in $L$. In that case, no repetition would actually be
observed in $L$, so that the derivation of a loop operator is not meaningful.
\item $P$ contains a concurrency operator only if the behavior of its children
indeed occurs in any order in the log. If only one of the orderings
appears in the log, then $P$ should include a sequence operator,
instead of the concurrency operator, to concisely capture the log behavior.

\end{compactitem}

\smallskip
\noindent

\textit{Maximality:} COBPAM ensures maximality of discovered patterns by construction: when a pattern is added to the set of frequent patterns, its seeds are deleted from the list.

Consider the trees \textit{seq(a,b)}, \textit{seq(a,c)}, and
\textit{seq(a,and(b,c))} in \autoref{fig:combinations}(b). If all of them are
frequent, the smaller trees are
not maximal, since they are included in \textit{seq(a,and(b,c))}.
In our example in \autoref{fig:runExample}, we discover that a blood
test (BT)
is followed by both, checkout (CO) and retrieval of belongings
(RB), i.e., \textit{seq(BT,and(CO,RB))}. Here, patterns \textit{seq(BT,CO)} and
\textit{seq(BT,RB)} do not yield further insights.

%% file: alignment_growth.tex
\section{Alignment Growth}
\label{sec:aligrow}

The main idea of the existing approaches to discover behavioral patterns from
an event log is to proceed incrementally, and to reuse results obtained for
smaller patterns in the exploration of larger pattern candidates. Based on the monotonicity properties for quality metrics, the discovery of patterns is spedup through pruning. However, each time the support of a pattern is assessed, the whole
respective process tree is aligned with a complete trace. The computation of such an alignment is computationally hard.

In this section, we extend the idea of reusing results from smaller patterns for the evaluation of each pattern candidate. 
In combination based pattern discovery (COBPAM), we show how to rely on the alignments computed already for the seeds of the pattern to
construct an alignment for the process tree at hand. We refer to this procedure
as \emph{alignment growth}.

Given a process tree and a trace, our alignment growth procedure recursively
accomplishes two tasks. First, it detects which
part of the behavior of the tree is already present in the trace and, hence,
has previously been aligned when handling the predecessors in the
construction tree. Second, it decides whether and which parts of the tree need
to be re-aligned with a certain projection of the trace.
This way, intuitively, any alignment that needs to be computed refers only
to a relatively small tree and a small part of the trace, thereby reducing the
runtime substantially.

Below, we first recall necessary definitions of alignments and
elaborate on an important property of alignment computation that is exploited
in our approach (\autoref{sec:alignments}). We then define the notion of
validated contexts (\autoref{sec:context}), before introducing the actual
alignment growth procedure (\autoref{sec:growthProc}). Finally, we exemplify
the procedure with a comprehensive example (\autoref{sec:example}).

\subsection{Alignments and the LOF Property}
\label{sec:alignments}

In order to evaluate the support of a pattern, according to
\autoref{def:support}, we need to assess whether a word of the language
$\Sigma(P)$ of the pattern $P$ is a subsequence of a trace $(id,\sigma)$. To
this end, one may construct an alignment between the activities of $\sigma$ and
the activities of a word $w\in \Sigma(P)$. Such an alignment comprises a
sequence of steps, each relating one activity of the trace to an activity of
the word, or an activity of either one to a placeholder to denote the absence
of a counterpart~\cite{DBLP:books/sp/CarmonaDSW18}. The steps are defined such
that, once placeholders are
removed, they yield the original trace or the original word, respectively.
For instance, consider a trace with the activity sequence $\sigma=\langle a, b,
c, b, b \rangle$ and the pattern $seq(a, and(b,c))$. Considering the word
$w=\langle a, c, b \rangle$ of the language of the pattern, two example
alignments are (where $\gg$ denotes a placeholder):

\smallskip
\begin{tabular}{l|c@{\hspace{0.7em}}c@{\hspace{0.7em}}
		c@{\hspace{0.7em}}c@{\hspace{0.7em}}c@{\hspace{0.7em}}
		c@{\hspace{0.7em}}
 		@{\hspace{2em}} l|
 		c@{\hspace{0.7em}}c@{\hspace{0.7em}}c@{\hspace{0.7em}}
 	    c@{\hspace{0.7em}}c@{\hspace{0.7em}}c@{\hspace{0.7em}}}
$w$ & $\boldsymbol{a}$ & $\gg$ & $\boldsymbol{c}$ & $\boldsymbol{b}$ & $\gg$ &
& $w$ &
$\boldsymbol{a}$ & ${c}$ & $\boldsymbol{b}$ & $\gg$ & $\gg$& $\gg$ \\
\cmidrule{1-6} \cmidrule{8-14}
$\sigma$ & $\boldsymbol{a}$& $b$& $\boldsymbol{c}$& $\boldsymbol{b}$& $b$ & &
$\sigma$ & $\boldsymbol{a}$& $\gg$ & $\boldsymbol{b}$ & $c$& $b$& $b$\\
\end{tabular}

\smallskip
In order to check whether a word $w\in \Sigma(P)$ of a pattern language is a
subsequence of a trace $(id,\sigma)$, only alignments that assign placeholders
for activities of the word $w$ are to be considered. Put differently, all
activities of the word $w$ need to occur in $\sigma$, whereas some activities
in $\sigma$ may not appear in $w$. From the above alignments, only the first
one satisfies this requirement. The second alignment, in turn, assigns a placeholder also to an activity of the trace.

From the above notion of an alignment, we derive a structure, coined
\emph{shadow map}, that maps the indices of a word of a pattern
language and the indices of its occurrence in the trace. This mapping preserves the order in
the word and the trace. We define this structure as
follows:

\begin{definition}[Shadow Map]
Let $\sigma'$ be a sequence of activities and let $(id,\sigma)$ be a trace,
such that $\sigma' \sqsubseteq \sigma$.
A shadow map of $\sigma'$ on $\sigma$ is any function $s:\{1, 2, \dots,
|\sigma'| \}
\rightarrow \{1, 2, \dots, |\sigma|\}$, where for all $i,j \in \mathbb{N}, 1
\leq i \leq |\sigma'|, 1 \leq j \leq |\sigma|$, it holds that $s(i) = j$ if $\sigma'(i)
= \sigma(j)$.

\end{definition}
We may write $s(\sigma'(i))$ to refer to $s(i)$, when $\sigma'(i)$ is
unique in the trace. Also, we call $s(i)$ the shadow of $\sigma'(i)$.
Considering the left of the above alignments, the shadow map is given as
$s(1)=1$, $s(2)=3$, and $s(3)=4$.

Given a trace and a word of a pattern language, there may exist multiple
alignments of the aforementioned type and, hence, multiple shadow maps may be
derived. In such a case, we assume that the one with the
smallest indices for each activity is chosen. We characterize the respective
shadow map by the leftmost-occurrence-first (LOF) property:

\begin{property}
	Let $\sigma'$ be a sequence of activities, let $(id,\sigma)$ be a trace
	with $\sigma' \sqsubseteq \sigma$, and let $s$ be a shadow map of $\sigma'$
	on
	$\sigma$. Then, $s$ shows the leftmost-occurrence-first (LOF) property,
	if for all $2 \leq i \leq |\sigma'|$ with $s(i-1) = j', s(i) = j$, it holds
	that there does not exist $j' < k < j$ with $\sigma'(i) = \sigma(k)$ and
	there does not exist $1 \leq k < s(1)$ with $\sigma'(1) = \sigma(k)$.
\end{property}
We note that several existing algorithms for the construction of alignments,
such as those based on $A^*$ search, yield alignments for which the derived
shadow maps satisfy the LOF-property. Moreover, our alignment growth procedure
maintains this property, meaning that any alignment constructed for a pattern
based on alignments of its seeds yields a shadow map with the LOF property, if
the alignments of the seeds showed the property already.

Turning to our example again, for the trace $\langle a, b, c, b, b \rangle$ and
the word $\langle a, c, b \rangle$ of the language of pattern $seq(a,
and(b,c))$, the above shadow map induced by
$\langle \boldsymbol{a}, {b}, \boldsymbol{c}, \boldsymbol{b}, {b}  \rangle$
satisfies the LOF property.

\begin{definition}[Boundary of a Pattern]
	Let $s$ be a shadow map of pattern $P$ on a trace $(id, \sigma)$.
	Then, the boundary of $P$ in $\sigma$ induced by $s$,
	symbolized by $\beta(\sigma, P, s)$ is the pair $(s(1),
	s(|\upsilon(\sigma, P)|))$. The lowest index $\beta_l(\sigma, P, s)$ is
	called lower boundary
	and the highest index $\beta_h(\sigma, P,s)$, the highest boundary. If the
	context
	is clear, we may write $\beta_l(P,s)$, $\beta_h(P,s)$ and $\beta(P,s)$.
\end{definition}

\subsection{Validated Context in Alignments}
\label{sec:context}

Our goal is to reuse alignments from previously evaluated trees. For that, we simply leverage the already computed shadow maps of the seeds. Given a pattern $P$ to evaluate on a trace \textit{(id, $\sigma$)}, we construct its shadow map $s$ and a fortiori align it by potentially incorporating parts from the older shadow maps. In other words, while we search for a word $w$ of the language of $P$ in $\sigma$, we detect the presence of subsequences of $w$ in the trace documented by previous shadow maps. With respect to $w$, these subsequences represent a context to which other activities can be added through concatenation and interleaving in order to construct the final word of the language. Since their presence is already validated through previous alignments, we call them Validated Context.

\begin{definition}
    Let a process tree $P$, its seeds $P_1$ and $P_2$ and their respective shadow maps $s$, $s_1$, $s_2$, With the objective of aligning $P$ on a trace $(id, \sigma)$, we call Validated Context a part of a word of the language of $P$ whose alignment on $\sigma$ can be deduced directly from alignments of $P_1$ and $P_2$. We note it $V(\sigma, P)$ or, if the context is clear, $V(P)$. $V(P)$ satisfies the following property involving the shadow maps: $ \forall c \in V(\sigma, P), s(c) = s_1(c) \vee s(c) = s_2(c)$. 
\end{definition}

\begin{figure}[h]
	\centering

	\subfloat[]{\centering
		\begin{tabular}{c}
			\begin{tikzpicture}
				\tikzset{level 1/.style={level distance=24pt}}
				\tikzset{level 2/.style={level distance=20pt}}
				\tikzset{level 3+/.style={level distance=20pt}}
				\Tree [.\textit{seq} [.\textit{loop} [.\textit{seq} [.b ] [.a ]
				] [.c ] ] [.\textit{and} [.d ] [.e ] ] ]
			\end{tikzpicture}
		\end{tabular}
 }\quad \quad
	\subfloat[]{\centering
		\begin{tabular}{c}
			\begin{tikzpicture}
				\tikzset{level 1/.style={level distance=24pt}}
				\tikzset{level 2/.style={level distance=20pt}}
				\tikzset{level 3+/.style={level distance=20pt}}
				\Tree [.\textit{seq} [.\textit{loop} [.a ] [.c ] ]
				[.\textit{and} [.d ] [.e ] ] ]
			\end{tikzpicture}
		\end{tabular}
 }\quad \quad 
	\subfloat[]{\centering
		\begin{tabular}{c}
			\begin{tikzpicture}
				\tikzset{level 1/.style={level distance=24pt}}
				\tikzset{level 2/.style={level distance=20pt}}
				\tikzset{level 3+/.style={level distance=20pt}}
				\Tree [.\textit{seq} [.\textit{loop} [.b ] [.c ] ]
				[.\textit{and} [.d ] [.e ] ] ]
			\end{tikzpicture}
		\end{tabular}
	}

	\vspace{-.5em}
	\caption{A tree $P$ (a) and its seeds: $P_1$ (b) and $P_2$ (c).}
	\label{fig:loopBlock}
	\vspace{-1em}
\end{figure}

\begin{algorithm}[t]
	\tiny
	\caption{Discovery}
	\label{alg:todoh}
	\SetKwInOut{Input}{input}%
	\SetKwInOut{Output}{output }%
	\Input{
		$L$, an event log; $\tau_f$, a support threshold;
	}
	\Output{
		$\mathcal{P}_{\mathit{freq}}$, a set of frequent maximal patterns}
	\SetKwFunction{align}{align}
	\SetKwFunction{eval}{eval}
	\BlankLine
	$A_L \gets  \bigcup_{(id,\langle a_1,\ldots, a_n\rangle )\ \in\ L, 1\leq
	j\leq n} \{a_j \}$\tcp*{Collect all actvities of traces of the log}
	$S\gets \{ x(a_1,a_2) \mid x\in \Omega \land a_1,a_2 \in A_L \land a_1\neq
	a_2 \}$\tcp*{Consider all process trees built of two activities as seeds,
	i.e., atomic patterns}
	\While{not terminate}{
		$S_{\mathit{freq}} \gets \{ P\in S \mid \mathit{support}(P, L) \geq
		\tau_f\}$\tcp*{Obtain frequent seeds}
		$S_{\mathit{infreq}} \gets S\setminus S_{\mathit{freq}}$\tcp*{Obtain
		infrequent seeds}
		\For{$P_1,P_2 \in S_{\mathit{freq}} \text{ with } P_1 \neq P_2 \land
		\mathit{combinable}(P_1,P_2)$}{
			$C \gets \{x(P_1, P_2) \mid x\in \{\mathit{and},\mathit{seq},
			\mathit{loop}\}\}$\tcp*{Construct candidate patterns of
			frequent seeds}
			$L' \gets \{(id,\sigma) \in L \mid \epsilon(\sigma, P_1) \cap
			\{(id,\sigma) \in L \mid \epsilon(\sigma, P_2)$\tcp*{TODO}

			\For{$P_3 \in C$}{
				$P_{loop} \gets L(a, P_3)$ with $a$ as the combination leaf of
				$P_1$\;
				$sup \gets 0$\;
				\For{$t\in L'$}{
					$map \gets \eval(P_3, t, P_1, P_2, P_{loop})$\;
					\lIf{$map \neq \emptyset$}{
						$sup \gets sup +1$
					}
				}
				\lIf{$sup \geq \tau_f$} {
					$S_{\mathit{freq}}\gets S_{\mathit{freq}} \cup \{P_3\}$
				}
				\lElse{
					$S_{\mathit{infreq}}\gets S_{\mathit{infreq}} \cup \{P_3\}$
				}

			}

		}
		\For{$P_1,P_2 \in S_{\mathit{infreq}} \text{ with } P_1 \neq P_2 \land
		\mathit{combinable}(P_1,P_2)$}{
			$P_3 \gets xor(P_1,P_2)$\;
			$sup \gets |\{(id,\sigma) \in L \mid \epsilon(\sigma, P_1) \cup
			\{(id,\sigma) \in L \mid \epsilon(\sigma, P_2)\}|$\;

			\If{$sup \geq \tau_f$} {
				$S_{\mathit{freq}}\gets S_{\mathit{freq}} \cup \{P_3\}$\;
                $P_{\mathit{freq}}\gets P_{\mathit{freq}} \setminus$ Construction tree of $P_3$\;
                $P_{\mathit{freq}}\gets P_{\mathit{freq}} \cup \{P_3\}$\;
                
			}
			\Else{
				$S_{\mathit{infreq}}\gets S_{\mathit{infreq}} \cup \{P_3\}$
			}
		}
	}

\end{algorithm}
\begin{algorithm}
	\tiny
	\caption{Alignment}
	\label{alg:alignment}
FUNCTION \eval{$P, \sigma, P_1, P_2,
\mathit{map}_1,\mathit{map}_2,
P_{loop}$}\;
	$\mathit{map} \gets \emptyset$\;
	\If(\tcp*[h]{Case 3}){$P = P_{loop}$}{
		\Return $map$, as obtained from traditional alignment algorithm of $P$
		on $\sigma$\;
	}
	\ElseIf(\tcp*[h]{Case 4}){$P = \mathit{seq}(a,b)$ with $a$ being
	combination leaf of $P_1$ and
	$b$ being combination leaf of $P_2$ }{
		\If{$\mathit{map}_1(a) < \mathit{map}_2(b)$} {
			$\mathit{map}(a)\gets \mathit{map}_1(a)$\;
			$\mathit{map}(b)\gets \mathit{map}_2(b)$\;
			\Return $\mathit{map}$\;
		}
		\Else{
			$\mathit{map}(a)\gets \mathit{map}_1(a)$\;
			\If{there exists $b$ after $\mathit{map}_1(a)$ in $\sigma$} {
				$\mathit{map}(b)\gets$ first occurrence of b after
				$\mathit{map}_1(a)$ in $\sigma$\;

				\Return $\mathit{map}$\;
			}
			\lElse {
				\Return $\emptyset$
			}
		}
	}
	\ElseIf(\tcp*[h]{Case 5}){$P = \mathit{and}(a,b)$ with $a$ being
		combination leaf of $P_1$ and
		$b$ being combination leaf of $P_2$}{
			$\mathit{map}(a)\gets \mathit{map}_1(a)$\;
			$\mathit{map}(b)\gets \mathit{map}_2(b)$\;
		\Return $\mathit{map}$\;
	}
	\ElseIf(\tcp*[h]{Case 6}){$P = \mathit{loop}(a,b)$ with $a$ being
	combination leaf of $P_1$ and
	$b$ being combination leaf of $P_2$}{
		\Return $map$, as obtained from traditional alignment algorithm of $P$
		on $\sigma$\;
	}
	\ElseIf{$P = \mathit{seq}(Q_1,Q_2)$}{
		\eIf(\tcp*[h]{Case 1.1}){$P_{loop}$ is a subtree of $Q_2$}{
		$\mathit{map} \gets \mathit{map} \cup \mathit{map}_1|_{alphabet(Q_1)}$\;
		$\sigma' \gets \mathit{tail}^{\beta_h(Q_1, map_1) + 1}(\sigma)$\;
		$\mathit{map}\gets \mathit{map} \cup \eval(Q_2, \sigma',$ right child
		of $root(P_1), $ right child of $ root(P_2), \mathit{map}_1|_{right
		child of P_1},
\mathit{map}_2|_{right child of P_2}, P_{loop})$\;
		\Return $\mathit{map}$\;
		}{
		$\mathit{map}\gets \eval(Q_1, \sigma,$ left child
		of $root(P_1), $ left child of $ root(P_2), \mathit{map}_1|_{left child
		of P_1},
		\mathit{map}_2|_{left child of P_2}, P_{loop})$\;
		\eIf{$\beta_h(Q_1, \mathit{map})< \beta_l( Q_2,
		\mathit{map}_1)
			\lor \beta_h(Q_1,\mathit{map} )< \beta_l(Q_2,
			\mathit{map}_2)$}{
			\eIf{$\beta_l(Q_2, \mathit{map}_1)< \beta_l(Q_2,
				\mathit{map}_2)$}{
			$\mathit{map} \gets \mathit{map} \cup
			\mathit{map}_1|_{alphabet(Q_2)}$
			}{
			$\mathit{map} \gets \mathit{map} \cup
			\mathit{map}_2|_{alphabet(Q_2)}$
			}
			\Return $map$\;
		}{
		\Return $map$, as obtained from traditional alignment algorithm of
		$Q_2$ on $\mathit{tail}^{\max\{\beta_l(Q_2,
		\mathit{map}_1), \beta_l(Q_2,
		\mathit{map}_2)\}+1}(\sigma)$\;
		}
		}
	}
	\ElseIf(\tcp*[h]{Case 2}){ $P = \mathit{and}(Q_1,Q_2)$}{
		\lIf{$P_{loop}$ is a subtree of $Q_2$}{
			swap $Q_1$ and $Q_2$
		}
		$\mathit{map} \gets \mathit{map}_1|_{alphabet(Q_2)}\cup
		\eval(Q_1, \sigma,$ left child
		of $root(P_1), $ left child of $ root(P_2), \mathit{map}_1|_{left child
			of P_1},
		\mathit{map}_2|_{left child of P_2}, P_{loop})$\;
		\Return $\mathit{map}$\;
	}

\end{algorithm}

\subsection{The Growth Procedure - or Alignments reuse in patterns combination operation}
\label{sec:growthProc}

For a new pattern P constructed out of the combination of two frequent seeds P1 and P2, and for each trace,  we determine the part of alignments of P1 or P2 that can be reused and the behaviors to realign.

When trying to reuse alignments for a new pattern, the loop operator and its position in the pattern play an important role.

\begin{definition}[The Loop Block]
	For a given leaf \emph{a} for a process tree $P$, the loop block of
	\emph{a} in $P$, called $L(a, P)$, is a subtree of $P$ whose root, noted
	\emph{r}, is an ancestor operator of \emph{a}. \emph{r} is a loop operator
	and no ancestor of \emph{r} is a loop operator. When no ancestor of
	\emph{a} is a loop operator or when $P$ is a leaf, $L(a, P ) = a$.
\end{definition}

The intuition behind the loop block and its role in the alignment reuse procedure can be illustrated using the following example. We
consider the pattern $P$ in \autoref{fig:loopBlock} and its two seeds $P_1$ and
$P_2$. Let the following trace $\sigma = \langle a, c, a, b, c, b, d, e
\rangle$ where both $P_1$ and $P_2$ appear in the respective forms: $\sigma =
\langle \mathbf{a}, \mathbf{c}, \mathbf{a}, b, c, b, \mathbf{d}, \mathbf{e}
\rangle$ and $\sigma = \langle a, c, a, \mathbf{b}, \mathbf{c}, \mathbf{b},
\mathbf{d}, \mathbf{e} \rangle$.

Now, if we want to align $P$, the left child of the loop becomes \textit{seq(b,
	a)}; meaning, we have to find the word $\langle b, a \rangle$, followed by
$c$
and then a repetition of $\langle b, a \rangle$. So, with respect to the seeds,
the combination operation brought two changes in what we are searching for, one
before the right child of the loop $c$ and another after; whereas the sequence
and the concurrence operators bring only one change. Both these alterations
need to be aligned and the validated context is minimal: $b$ (since we are searching for $\langle b, a, c, b, a \rangle$, we take the first $b$ in the trace as a validated context). Indeed, we search
for $\langle a, c, b, a \rangle$ in $\sigma = \langle a, c, a, \mathbf{b},
\underrightarrow{c, b, \mathbf{d}, \mathbf{e}} \rangle$ which doesn't exist in
this case ($\underrightarrow{}$ indicates graphically the tail of the trace after the validated context). In the end, most of the behavior we are searching for needs a
complete alignment and detecting a validated context comes down to assessing a
lot of individual cases with a low probability of seeing actual gains. An
extreme example is having a cascade of loops like : $P' =
loop(loop(loop(seq(a,b), c), d), e)$. When we align this tree, the change
introduced is only at the level of $seq(a,b)$ but it has a repercussion on all
the higher loops. Each one of them has to modify the first occurrence of the
left child as well as its repetition. A complete realignment of the highest
loop containing the change is necessary; which is exactly the definition of the
loop block.

In the following we will explain the alignment growth procedure.  Having at disposal a new pattern $P$ constructed out of the combination of two frequent seeds $P_1$ and $P_2$ (according to the monotonicity property), the
objective is to determine for each trace, the validated context and the
behaviors to realign. Of course, in the case where $P$ is a leaf, a simple
classical alignment is applied (in reality, it is a simple reading of the trace
from left to right until the leaf is found). If not, we set $a$ as the
combination leaf in $P_1$ and $b$, the combination leaf in $P_2$.

We need to precise that the trees containing a choice operator are aligned classically. The reason will be specified later. An exception is the
trees combined through the \textit{xor} operator whose frequency can be inferred  from the frequencies of their seeds and thus do not require an alignment (see~\cite{acheli2019efficient}).

Let $(id, \sigma)$ a trace and $map_1$ (resp. $map_2$) the
shadow map of $P_1$ on $\sigma$ (resp. $P_2)$. We execute the following
recursive algorithm. We bring
attention on the fact that the alignment issued from this algorithm respects
the LOF property which is proved recursively. Moreover, any alignment of $P$
is executed on $\sigma \uparrow \mathcal{A}(P)$ since the other activities
can't be in the alignment result anyway. For simplification, this is implied
when not mentioned.

The objective is to align a tree $P$ using the growth method on a trace tail
$\sigma$ constructing in the process its shadow map : $map$. In other terms, we are searching for a validated context while
replaying some parts of $P$. We suppose $P$ is constructed out of the
combination of two seeds $P_1$ and $P_2$ where $a$ and $b$ are the respective
combination leaves. $P_1$ and $P_2$ have thus already been aligned. We face
the following cases:

\textbf{Case 1:} if $P_1$ is not a leaf and $P$ writes as $seq(Q_1, Q_2)$ with
$Q_1$, $Q_2$ two subtrees. Then, we will enumerate two cases:

~\textbf{Case 1.1:} if $L(a, P)$ is contained in $Q_2$, then according to
the workings of the combination operation, $P_1$ writes as, $P_1 = seq(Q_1,
Q_{21})$ and $P_2 = seq(Q_1, Q_{22})$ with $Q_{21}, Q_{22}$, the two seeds of
$Q_2$ and the right children of the roots of $P_1$ and $P_2$ respectively. As such, $Q_1$ was already aligned on $\sigma$ while respecting the LOF
property: the occurrence computed is the leftmost one. This implies that $Q_1$
has the same shadow map in both seeds. Consequently, all we need to do is to
align $Q_2$ using the alignment growth algorithm after $Q_1$, meaning on
$\mathit{tail}^{\beta_h(Q_1, map_1) + 1}(\sigma)$. Indeed, since the occurrence of $Q_1$
considered is the leftmost one, any occurrence of $Q_2$ that validates $seq(Q_1,
Q_2)$ comes after $Q_1$. Moreover, since both alignments respect the LOF
property then the whole alignment of $P$ respects it too. It is to be noted
that the shadow map of $Q_1$ on $\sigma$, extracted as $\mathit{map}_1|_{alphabet(Q_1)}$, is added directly to $map$ as it is a part of the validated context of
$P$ and that the seeds of $Q_2$, $Q_{21}$ and $Q_{22}$, were aligned on
$\mathit{tail}^{\beta_h(Q_1, map_1) + 1}(\sigma)$.

Finally, We set $P$ to $Q_2$ and $\sigma$ to $\mathit{tail}^{\beta_h(Q_1, map_1) +
1}(\sigma)$
and call the alignment growth algorithm recursively.

\textbf{Case 1.2:} if $L(a, P)$ is contained in $Q_1$, then, according to the
combination operation, $P_1 = seq(Q_{11}, Q_2)$ and $P_2 = seq(Q_{12}, Q_2)$ with
$Q_{11}$, $Q_{12}$ the seeds of $Q_1$. This means that $Q_2$ was aligned in both
seeds. However, they don't necessarily share the same shadow map, as in $P_1$,
$Q_2$ was aligned after $Q_{11}$ and in $P_2$ after $Q_{12}$. We will note
$\beta^1 = \beta(Q_2, map_1)$ the boundary of the appearance of $Q_2$ during the alignment of
$P_1$ and $\beta^2 = \beta(Q_2, map_2)$ that of its appearance during the alignment of $P_2.$ On
the same subject, while it is true that $Q_2$ is aligned in $\sigma$, we have
no guarantee that $Q_1$ exists in $\sigma$ before the already calculated
appearances of $Q_2$ (either before $\beta^1_l$ or $\beta^2_l$). So we have to
align $Q_1$ using the growth algorithm on $\sigma$ (we set $P$ to $Q_1$ and
enter a new recursion. $\sigma$ stays unchanged). Once that done, we
check if $\beta_h(Q_1, map) < min(\beta^1_l, \beta^2_l)$ and if not, we check
$\beta_h(Q_1, map) < max(\beta^1_l, \beta^2_l)$. If the first condition is met, then
the leftmost occurrence of $Q_2$ calculated in the seeds and recorded in $map_1$ is the one aggregated
into the validated context. Else, it is the second leftmost occurrence
calculated in the seeds and recorded in $map_2$. That is because in those cases, $Q_1$ still precedes
$Q_2$ in $\sigma$. If neither conditions are met, we realign classically
$Q_2$ on the trace $\mathit{tail}^{\beta_h(Q_1, map) + 1}(\sigma)$. On another note,
since both
the alignment of $Q_1$ and $Q_2$ adhere to the LOF property, the whole alignment
does.

\textbf{Case 2:} if $P_1$ is not a leaf and $P$ writes as $P = and(Q_1, Q_2)$
with $Q_1$, $Q_2$ two subtrees, we test if $Q_2$ contains $L(a, P)$. In that
case, we swap $Q_1$ and $Q_2$ since the operator is symmetric and that has
neither an impact on the language of the tree nor on the alignment. As a
result, we have $P_1 = and(Q_{11}, Q_2)$ and $P_2 = and(Q_{12}, Q_2)$ with
$Q_{11}$, $Q_{12}$, the two seeds of $Q_1$. Since the seeds $P_1$ and $P_2$ were already aligned
then $Q_2$ was too. Besides, when aligning $P$, there is no order constraint
between $Q_2$ and $Q_1$. As such, the already calculated alignment of $Q_2$
serves as a validated context for $P$. The next step is to align $Q_1$ using
the growth algorithm on $\sigma$. The trace we align on doesn't change
because of the absence of an order enforcement. So we set only $P$ to $Q_1$ before the recursive call.
Finally, the alignment of $P$ respects the LOF property because both the
alignments of $Q_1$ and $Q_2$ do.

\textbf{Case 3:} $P_1$ is not a leaf and $P$ is $L(a, P)$. In this case, a
classical alignment of $P$ is required. There is no validated context $V(P)$
and the LOF property is respected thanks to the classical algorithm.

\textbf{Case 4:} $P = seq(a,b)$. We know that $a$ and $b$ have already been
aligned. So, we test if $map_2(b) > map_1(a)$. In that case, no classical
alignment is needed and the shadow maps of $a$ and $b$ are part of the
validated context. Else, $b$ is realigned classically after $a$. In other
words, $b$ is re-aligned on $\mathit{tail}^{map_1(a)+1}(\sigma)$. In both
cases, the
LOF property is respected because we use the classical procedure right after a
leftmost first occurrence and/or use already calculated alignments.

\textbf{Case 5:} $P = and(a,b)$. We know that $a$ and $b$ are aligned. Since
there is no order constraint between them then $P$ is aligned too. No
classical alignment is needed and the shadow maps of both $a$ and $b$ are
included in $V(P)$. In this particular case, the whole alignment of $P$ is
directly constructed and $\upsilon(\sigma, P)$ is a validated context in its whole. Here too, the LOF
property is satisfied.

\textbf{Case 6:} $P = loop(a,b)$. There is no validated context and $P$ is
aligned classically satisfying thus the LOF property.

The reason to exclude the \textit{xor} operator from the procedure is that  it puts us in a position similar to that of the loop, where the existing alignments may not be very helpful. Indeed, for the \textit{xor} operator, the alignment of a seed may involve a \textit{xor} branch, while an alignment for the new pattern would necessarily concern other \textit{xor} branches (no alignment is possible through the old branch). The following example illustrates this.

Let $P'_1 = seq(xor(a, b), d)$, $P'_2 = seq(xor(a, c), d)$ and a trace
$\sigma' = \langle c, b, a, d \rangle$. $P'_1$ and $P'_2$ appear respectively
as: $\langle c, \mathbf{b}, a, \mathbf{d} \rangle$ and $ \langle \mathbf{c}, b,
a, \mathbf{d} \rangle$. Our objective is to align $P' = seq(xor(a, seq (b, c)),
d)$ that introduces a change under the xor operator with a new seq operator. In the growth algorithm, we tend to realign the changes,
while mostly leaving the leftmost validated context untouched. However,
aligning the changes below the choice operator is not the right thing to do.
Indeed, the realignment would take the form $ \langle c, \mathbf{b},
\underrightarrow{a, d} \rangle$ and wouldn't yield a result. Conversely,
adopting an other path in the $xor$ subtree allows us to determine an
appearance of $P$ : $\langle c, b, \mathbf{a}, \mathbf{d} \rangle$. This path
will bypass the new changes that are non existent in the trace. It is important
to know that the $a$ path was not chosen initially because the other path was
encountered earlier in the trace.

As the handling of the $xor$ operator for alignment reuse is case-dependent leading to a complex algorithm with many specific cases and
with limited reuse, we decided to align classically trees containing \textit{xor} operators.

On a side note, a hidden feature that is not directly visible is that when excluding the
validated context from the alignment, we end up excluding its alphabet from the
trace. As we mentioned earlier, the realignment of any behavior is realized on
a version of the trace containing only the relevant activities. So, in addition
to truncating the trace and using only its tail, the number of activities to
align on is reduced. The classical alignment on the other hand proceeds using
the alphabet of the entire tree which is less than optimal.

\subsection{An Illustrative Example}
\label{sec:example}

We present in the following an instance of an alignment using the growth
procedure. The tree to align is $R$ depicted in \autoref{fig:exAlign} along
with its seeds $R_1$ and $R_2$. We suppose the trace to align on is: $\sigma =
\langle a, e, f, c, b, c, a, b, c, d, f, e  \rangle$. The seed $R_1$ is
contained as: $\langle \mathbf{a}, e, f, \mathbf{c}, b, c, a, b, c, \mathbf{d},
\mathbf{f}, \mathbf{e} \rangle$ and $R_2$ as: $\langle a, e, f, c, \mathbf{b},
\mathbf{c}, a, b, c, \mathbf{d}, \mathbf{f}, \mathbf{e} \rangle$.  Here's the
alignment's recursive evolution. We first set $P = R$. $\sigma$ is unchanged.

\begin{enumerate}
	\item \textbf{(Case 1.2.)} We have $P_1 = R_1$ is not a leaf and $P = R$
	writes as a sequence. As $L(a, R) = a$ is contained in the left child,
	we realign \textit{and(seq(seq(a, b), c), d)} on $\sigma$. We set $P =
	and(seq(seq(a, b), c), d)$, $P_1 = and(seq(a, c), d)$ and $P_2 =
	and(seq(b, c), d)$ and leave $\sigma$ unchanged.
	\begin{enumerate}
		\item \textbf{(Case 2.)} We have $P_1$ is not a leaf and the root of
		$P$ is a concurrence operator. As such, $d$ is a validated context.
		The shadow map of $P$ and a fortiori of $R$ contains $\langle a, e, f,
		c, b, c, a, b, c, \mathbf{d}, f, e \rangle$. We set $P = seq(seq(b,
		a), c)$ with $P_1 = seq(a, c)$, $P_2 = seq(b, c)$ and leave $\sigma$ unchanged.
		\begin{enumerate}
			\item \textbf{(Case 1.2.)} We have $L(a, P) = a$ contained in
			the left child of $P$. So we have to align the first child $P_1$. We set $ P = P_1 = seq(b, a)$ and leave $\sigma$ unchanged. In this case, the new values of $P_1$ and $P_2$ are $P_1 = a$ and
			$P_2 = b$.
			\begin{enumerate}
				\item \textbf{(Case 4.)} We have $P_1$ is a leaf and $P =
				seq(b, a)$. $b$ and $a$ have already been aligned in the forms:
				$\langle \mathbf{a}, e, f, c, b, c, a, b, c, d, f, e \rangle$
				and\\ $\langle a, e, f, c, \mathbf{b}, c, a, b, c, d, f, e
				\rangle$ respectively. However, we notice that in this
				alignment $a$ is not after $b$. So we have to align $a$ after
				$b$:\\ $\langle a, e, f, c, \mathbf{b}, \underrightarrow{c, a,
				b, c, d, f, e} \rangle$; (the actual trace aligned is the
				projection on the considered alphabet: $\langle a, e, f, c,
				\mathbf{b}, \underrightarrow{a} \rangle$) which results in\\
				$\langle a, e, f, c, \mathbf{b}, c, \mathbf{a}, b, c, d, f, e
				\rangle$. That is the alignment of \textit{seq(b, a)}.
			\end{enumerate}
			\item (Follow-up of (i), \textbf{Case 1.2.}) The left child
			\textit{seq(a, b)} of $P$ has just been aligned. We have that
			$P_1 = seq(a, c)$ and $P_2 = seq(b, c)$ have already been aligned
			in the seeds in the respective forms: $\langle \mathbf{a}, e, f,
			\mathbf{c}, b, c, a, b, c, d, f, e \rangle$ and $\langle a, e, f,
			c, \mathbf{b}, \mathbf{c}, a, b, c, d, f, e \rangle$. Both the
			occurrences of $c$ in these two forms are before the occurrence of
			$seq(b, a)$. So we have to realign $c$ after the latter occurrence:
			$\langle a, e, f, c, \mathbf{b}, c, \mathbf{a}, \underrightarrow{b,
			c, d, f, e} \rangle$ (here too, the actual trace realigned:
			$\langle a, e, f, c, \mathbf{b}, c, \mathbf{a}, \underrightarrow{c}
			\rangle$) which results in $\langle a, e, f, c, \mathbf{b}, c,
			\mathbf{a}, b, \mathbf{c}, d, f, e \rangle$. This is the alignment
			of $P = seq(seq(b, a), c)$.
		\end{enumerate}
		\item (Follow-up of (a), \textbf{Case 2.}). The only thing to do here
		is to incorporate the validated context $d$ and construct the alignment
		of $P$: $\langle a, e, f, c, \mathbf{b}, c, \mathbf{a}, b, \mathbf{c},
		\mathbf{d}, f, e \rangle$
	\end{enumerate}
	\item (Follow-up of (1), \textbf{Case 1.2.}) The left child of $R$:
	$and(seq(seq(a, b), c), d)$ has been aligned. Now we test if the already
	computed occurrence of the right child \textit{and(f,e)} exists after the
	first child. That is the case and the occurrence is considered a validated
	context. Finally, the complete alignment of $R$ is $\langle a, e, f, c,
	\mathbf{b}, c, \mathbf{a}, b, \mathbf{c}, \mathbf{d}, \mathbf{f},
	\mathbf{e} \rangle$.
\end{enumerate}

\begin{figure}[h]
	\centering

	\subfloat[]{\centering
		\begin{tabular}{c}
			\begin{tikzpicture}
				\tikzset{level 1/.style={level distance=24pt}}
				\tikzset{level 2/.style={level distance=20pt}}
				\tikzset{level 3+/.style={level distance=20pt}}
				\Tree [.\textit{seq} [.\textit{and} [.\textit{seq}
				[.\textit{seq} [.b ] [.a ] ] [.c ] ] [.d ] ] [.\textit{and} [.e
				] [.f ] ] ]
			\end{tikzpicture}
		\end{tabular}
	} \quad \quad 
	\subfloat[]{\centering
		\begin{tabular}{c}
			\begin{tikzpicture}
				\tikzset{level 1/.style={level distance=24pt}}
				\tikzset{level 2/.style={level distance=20pt}}
				\tikzset{level 3+/.style={level distance=20pt}}
				\Tree [.\textit{seq} [.\textit{and} [.\textit{seq} [.a  ] [.c ]
				] [.d ] ] [.\textit{and} [.e ] [.f ] ] ]
			\end{tikzpicture}
		\end{tabular}
	}\quad \quad \quad
	\subfloat[]{\centering
		\begin{tabular}{c}
			\begin{tikzpicture}
				\tikzset{level 1/.style={level distance=24pt}}
				\tikzset{level 2/.style={level distance=20pt}}
				\tikzset{level 3+/.style={level distance=20pt}}
				\Tree [.\textit{seq} [.\textit{and} [.\textit{seq} [.b ] [.c ]
				] [.d ] ] [.\textit{and} [.e ] [.f ] ] ]
			\end{tikzpicture}
		\end{tabular}
	}

	\vspace{-.5em}
	\caption{A tree $R$ (a) and its seeds: $R_1$ (b) and $R_2$ (c).}
	\label{fig:exAlign}
	\vspace{-1em}
\end{figure}

In conclusion, we took advantage of the validated contexts and aligned just two
times a single leaf on a single leaf trace. We reduced an exponentially hard
alignment problem to two \textit{if conditions}. In fact, the more the
validated context alignment is time consuming, the higher the gains.

%% file: paper-elsarticle.bbl
\begin{thebibliography}{10}
\expandafter\ifx\csname url\endcsname\relax
  \def\url#1{\texttt{#1}}\fi
\expandafter\ifx\csname urlprefix\endcsname\relax\def\urlprefix{URL }\fi
\expandafter\ifx\csname href\endcsname\relax
  \def\href#1#2{#2} \def\path#1{#1}\fi

\bibitem{VanderAalst2016ProcessAction}
W.~Van~der Aalst, {Process mining: Data science in action}, Springer Berlin
  Heidelberg, Berlin, Heidelberg, 2016.

\bibitem{DBLP:books/sp/DumasRMR18}
M.~Dumas, M.~L. Rosa, J.~Mendling, H.~A. Reijers,
  \href{https://doi.org/10.1007/978-3-662-56509-4}{Fundamentals of Business
  Process Management, Second Edition}, Springer, 2018.
\newblock \href {https://doi.org/10.1007/978-3-662-56509-4}
  {\path{doi:10.1007/978-3-662-56509-4}}.
\newline\urlprefix\url{https://doi.org/10.1007/978-3-662-56509-4}

\bibitem{DBLP:journals/tkde/AugustoCDRMMMS19}
A.~Augusto, R.~Conforti, M.~Dumas, M.~L. Rosa, F.~M. Maggi, A.~Marrella,
  M.~Mecella, A.~Soo,
  \href{https://doi.org/10.1109/TKDE.2018.2841877}{Automated discovery of
  process models from event logs: Review and benchmark}, {IEEE} Trans. Knowl.
  Data Eng. 31~(4) (2019) 686--705.
\newblock \href {https://doi.org/10.1109/TKDE.2018.2841877}
  {\path{doi:10.1109/TKDE.2018.2841877}}.
\newline\urlprefix\url{https://doi.org/10.1109/TKDE.2018.2841877}

\bibitem{vanderAalst2004WorkflowLogs}
W.~van~der Aalst, T.~Weijters, L.~Maruster, {Workflow mining: discovering
  process models from event logs}, IEEE Transactions on Knowledge and Data
  Engineering 16~(9) (2004) 1128--1142.

\bibitem{Burattin2012HeuristicsData}
A.~Burattin, A.~Sperduti, W.~M. van~der Aalst, Heuristics miners for streaming
  event data, arXiv preprint arXiv:1212.6383 (2012).

\bibitem{Leemans2013LNCSApproach}
S.~J. Leemans, D.~Fahland, W.~M. van~der Aalst, Discovering block-structured
  process models from event logs-a constructive approach, in: International
  conference on applications and theory of Petri nets and concurrency,
  Springer, 2013, pp. 311--329.

\bibitem{DBLP:conf/icdm/AugustoCDR17}
A.~Augusto, R.~Conforti, M.~Dumas, M.~L. Rosa,
  \href{https://doi.org/10.1109/ICDM.2017.9}{Split miner: Discovering accurate
  and simple business process models from event logs}, in: V.~Raghavan,
  S.~Aluru, G.~Karypis, L.~Miele, X.~Wu (Eds.), 2017 {IEEE} International
  Conference on Data Mining, {ICDM} 2017, New Orleans, LA, USA, November 18-21,
  2017, {IEEE} Computer Society, 2017, pp. 1--10.
\newblock \href {https://doi.org/10.1109/ICDM.2017.9}
  {\path{doi:10.1109/ICDM.2017.9}}.
\newline\urlprefix\url{https://doi.org/10.1109/ICDM.2017.9}

\bibitem{DBLP:journals/dss/BrouckeW17}
S.~K. L.~M. vanden Broucke, J.~D. Weerdt,
  \href{https://doi.org/10.1016/j.dss.2017.04.005}{Fodina: {A} robust and
  flexible heuristic process discovery technique}, Decis. Support Syst. 100
  (2017) 109--118.
\newblock \href {https://doi.org/10.1016/j.dss.2017.04.005}
  {\path{doi:10.1016/j.dss.2017.04.005}}.
\newline\urlprefix\url{https://doi.org/10.1016/j.dss.2017.04.005}

\bibitem{DBLP:conf/bpm/BoseA10}
R.~P. J.~C. Bose, W.~M.~P. van~der Aalst,
  \href{https://doi.org/10.1007/978-3-642-15618-2\_17}{Trace alignment in
  process mining: Opportunities for process diagnostics}, in: R.~Hull,
  J.~Mendling, S.~Tai (Eds.), Business Process Management - 8th International
  Conference, {BPM} 2010, Hoboken, NJ, USA, September 13-16, 2010. Proceedings,
  Vol. 6336 of Lecture Notes in Computer Science, Springer, 2010, pp. 227--242.
\newblock \href {https://doi.org/10.1007/978-3-642-15618-2\_17}
  {\path{doi:10.1007/978-3-642-15618-2\_17}}.
\newline\urlprefix\url{https://doi.org/10.1007/978-3-642-15618-2\_17}

\bibitem{DBLP:conf/icsoc/SunBW17}
Y.~Sun, B.~Bauer, M.~Weidlich,
  \href{https://doi.org/10.1007/978-3-319-69035-3\_12}{Compound trace
  clustering to generate accurate and simple sub-process models}, in: E.~M.
  Maximilien, A.~Vallecillo, J.~Wang, M.~Oriol (Eds.), Service-Oriented
  Computing - 15th International Conference, {ICSOC} 2017, Malaga, Spain,
  November 13-16, 2017, Proceedings, Vol. 10601 of Lecture Notes in Computer
  Science, Springer, 2017, pp. 175--190.
\newblock \href {https://doi.org/10.1007/978-3-319-69035-3\_12}
  {\path{doi:10.1007/978-3-319-69035-3\_12}}.
\newline\urlprefix\url{https://doi.org/10.1007/978-3-319-69035-3\_12}

\bibitem{DBLP:journals/kais/KoninckNBBSW21}
P.~D. Koninck, K.~Nelissen, S.~vanden Broucke, B.~Baesens, M.~Snoeck, J.~D.
  Weerdt, \href{https://doi.org/10.1007/s10115-021-01548-6}{Expert-driven trace
  clustering with instance-level constraints}, Knowl. Inf. Syst. 63~(5) (2021)
  1197--1220.
\newblock \href {https://doi.org/10.1007/s10115-021-01548-6}
  {\path{doi:10.1007/s10115-021-01548-6}}.
\newline\urlprefix\url{https://doi.org/10.1007/s10115-021-01548-6}

\bibitem{Maggi2011User-guidedModels}
F.~M. Maggi, A.~J. Mooij, W.~M. Van Der~Aalst, {User-guided discovery of
  declarative process models}, in: CIDM 2011, IEEE, 2011, pp. 192--199.

\bibitem{DiCiccio2016}
C.~{Di Ciccio}, F.~M. Maggi, J.~Mendling, {Efficient discovery of
  Target-Branched Declare constraints}, Information Systems 56 (2016) 258--283.
\newblock \href {https://doi.org/10.1016/j.is.2015.06.009}
  {\path{doi:10.1016/j.is.2015.06.009}}.

\bibitem{Tax2016MiningModels}
N.~Tax, N.~Sidorova, R.~Haakma, W.~M. van~der Aalst, {Mining local process
  models}, Journal of Innovation in Digital Ecosystems 3~(2) (2016) 183--196.

\bibitem{acheli2021discovering}
M.~Acheli, D.~Grigori, M.~Weidlich, Discovering and analyzing contextual
  behavioral patterns from event logs, IEEE Transactions on Knowledge and Data
  Engineering (2021).

\bibitem{acheli2019efficient}
M.~Acheli, D.~Grigori, M.~Weidlich, Efficient discovery of compact maximal
  behavioral patterns from event logs, in: International Conference on Advanced
  Information Systems Engineering, Springer, 2019, pp. 579--594.

\bibitem{Peeva21}
V.~Peeva, L.~L. Mannel, W.~van~der Aalst, Local process model discovery by
  combining places, in: Workshops of the International Conference on Process
  Mining, Springer, 2021.

\bibitem{Adriansyah2014AligningBehavior}
A.~Adriansyah, {Aligning Observed and Modeled Behavior}, Ph.D. thesis (2014).

\bibitem{DBLP:books/sp/CarmonaDSW18}
J.~Carmona, B.~F. van Dongen, A.~Solti, M.~Weidlich,
  \href{https://doi.org/10.1007/978-3-319-99414-7}{Conformance Checking -
  Relating Processes and Models}, Springer, 2018.
\newblock \href {https://doi.org/10.1007/978-3-319-99414-7}
  {\path{doi:10.1007/978-3-319-99414-7}}.
\newline\urlprefix\url{https://doi.org/10.1007/978-3-319-99414-7}

\bibitem{DBLP:journals/is/TaxDSAN18}
N.~Tax, B.~Dalmas, N.~Sidorova, W.~M.~P. van~der Aalst, S.~Norre,
  \href{https://doi.org/10.1016/j.is.2018.04.006}{Interest-driven discovery of
  local process models}, Inf. Syst. 77 (2018) 105--117.
\newblock \href {https://doi.org/10.1016/j.is.2018.04.006}
  {\path{doi:10.1016/j.is.2018.04.006}}.
\newline\urlprefix\url{https://doi.org/10.1016/j.is.2018.04.006}

\bibitem{Fournier-Viger2017AMining}
P.~Fournier-Viger, J.~Chun, W.~Lin, R.~U. Kiran, Y.~S. Koh, R.~Thomas, {A
  Survey of Sequential Pattern Mining}, Ubiquitous International 1~(1) (2017)
  54--77.

\bibitem{Srikant1996MiningImprovements}
R.~Srikant, R.~Agrawal, Mining sequential patterns: Generalizations and
  performance improvements, in: International Conference on Extending Database
  Technology, Springer, 1996, pp. 1--17.

\bibitem{Pei2004MiningApproach}
J.~Pei, J.~Han, B.~Mortazavi-Asl, J.~Wang, H.~Pinto, Q.~Chen, U.~Dayal, M.~C.
  Hsu, {Mining sequential patterns by pattern-growth: The prefixspan approach},
  IEEE Transactions on Knowledge and Data Engineering (2004).

\bibitem{DBLP:conf/icpm/PijnenborgVFLG21}
P.~Pijnenborg, R.~Verhoeven, M.~Firat, H.~van Laarhoven, L.~Genga,
  \href{https://doi.org/10.1109/ICPM53251.2021.9576880}{Towards evidence-based
  analysis of palliative treatments for stomach and esophageal cancer patients:
  a process mining approach}, in: C.~D. Ciccio, C.~D. Francescomarino,
  P.~Soffer (Eds.), 3rd International Conference on Process Mining, {ICPM}
  2021, Eindhoven, Netherlands, October 31 - Nov. 4, 2021, {IEEE}, 2021, pp.
  136--143.
\newblock \href {https://doi.org/10.1109/ICPM53251.2021.9576880}
  {\path{doi:10.1109/ICPM53251.2021.9576880}}.
\newline\urlprefix\url{https://doi.org/10.1109/ICPM53251.2021.9576880}

\bibitem{InteractiveLPM}
M.~Vazifehdoostirani, L.~Genga, X.~Lu, R.~Verhoeven, H.~van Laarhoven, R.~M.
  Dijkman, \href{https://doi.org/10.1007/978-3-031-41620-0\_18}{Interactive
  multi-interest process pattern discovery}, in: C.~D. Francescomarino,
  A.~Burattin, C.~Janiesch, S.~Sadiq (Eds.), Business Process Management - 21st
  International Conference, {BPM} 2023, Utrecht, The Netherlands, September
  11-15, 2023, Proceedings, Vol. 14159 of Lecture Notes in Computer Science,
  Springer, 2023, pp. 303--319.
\newblock \href {https://doi.org/10.1007/978-3-031-41620-0\_18}
  {\path{doi:10.1007/978-3-031-41620-0\_18}}.
\newline\urlprefix\url{https://doi.org/10.1007/978-3-031-41620-0\_18}

\bibitem{TaxThesis}
N.~Tax, \href{http://arxiv.org/abs/1909.01421}{Mining insights from
  weakly-structured event data}, CoRR abs/1909.01421 (2019).
\newblock \href {http://arxiv.org/abs/1909.01421} {\path{arXiv:1909.01421}}.
\newline\urlprefix\url{http://arxiv.org/abs/1909.01421}

\bibitem{de2013aligning}
M.~De~Leoni, W.~M. Van Der~Aalst, Aligning event logs and process models for
  multi-perspective conformance checking: An approach based on integer linear
  programming, in: Business Process Management, Springer, 2013, pp. 113--129.

\bibitem{koorneef2017automatic}
M.~Koorneef, A.~Solti, H.~Leopold, H.~A. Reijers, Automatic root cause
  identification using most probable alignments, in: International Conference
  on Business Process Management, Springer, 2017, pp. 204--215.

\bibitem{burattin2017framework}
A.~Burattin, J.~Carmona, A framework for online conformance checking, in:
  International Conference on Business Process Management, Springer, 2017, pp.
  165--177.

\bibitem{munoz2014single}
J.~Munoz-Gama, J.~Carmona, W.~M. Van Der~Aalst, Single-entry single-exit
  decomposed conformance checking, Information Systems 46 (2014) 102--122.

\bibitem{song2016efficient}
W.~Song, X.~Xia, H.-A. Jacobsen, P.~Zhang, H.~Hu, Efficient alignment between
  event logs and process models, IEEE Transactions on Services Computing 10~(1)
  (2016) 136--149.

\bibitem{verbeek2014decomposed}
H.~Verbeek, W.~M. van~der Aalst, Decomposed process mining: The ilp case, in:
  International conference on business process management, Springer, 2014, pp.
  264--276.

\bibitem{van2020repairing}
S.~J. van Zelst, J.~C. Buijs, B.~V{\'a}zquez-Barreiros, M.~Lama, M.~Mucientes,
  Repairing alignments of process models, Business \& Information Systems
  Engineering 62~(4) (2020) 289--304.

\bibitem{Buijs2012ATrees}
J.~C. Buijs, B.~F. Van~Dongen, W.~M. Van Der~Aalst, {A genetic algorithm for
  discovering process trees}, in: CEC 2012, IEEE, 2012, pp. 1--8.

\bibitem{DBLP:phd/hal/Acheli21}
M.~Acheli, \href{https://tel.archives-ouvertes.fr/tel-03542389}{Behavioral
  pattern mining for flexible processes. (fouille de patterns comportementaux
  dans le contexte de processus flexibles)}, Ph.D. thesis, {PSL} University,
  Paris, France (2021).
\newline\urlprefix\url{https://tel.archives-ouvertes.fr/tel-03542389}

\bibitem{vanDongen2005TheSupport}
B.~F. Van~Dongen, A.~K.~A. de~Medeiros, H.~Verbeek, A.~Weijters, W.~M. van
  Der~Aalst, The prom framework: A new era in process mining tool support, in:
  International conference on application and theory of petri nets, Springer,
  2005, pp. 444--454.

\end{thebibliography}
